\documentclass[aip,jcp,reprint,groupedaddress,floatfix]{revtex4-1}

\usepackage{algorithm}
\usepackage{algpseudocode}
\usepackage{amssymb}
\usepackage{dsfont}
\usepackage{graphicx}
\usepackage{mathtools}
\usepackage[version=4]{mhchem}
\usepackage{microtype}
\usepackage{physics}
\usepackage{relsize}
\usepackage{siunitx}

\usepackage[hidelinks]{hyperref}

\DeclareMathOperator*{\argmin}{argmin}

\DeclareMathOperator*{\erfcinv}{erfcinv}

\newcommand{\rem}{\mathbin{\mathsmaller{\scriptstyle \%}}}

\renewcommand{\vec}[1]{\boldsymbol{\mathbf{#1}}}

\renewcommand{\epsilon}{\varepsilon}
\renewcommand{\rho}{\varrho}
\renewcommand{\phi}{\varphi}

\newcommand{\midrule}{\hline \\[-3mm]}

\begin{document}

\title{
	Deterministic and quasi-random sampling of optimized Gaussian mixture distributions for vibronic Monte Carlo
}

\author{Dmitri Iouchtchenko}
\email{diouchtc@uwaterloo.ca}
\author{Neil Raymond}
\author{Pierre-Nicholas Roy}
\author{Marcel Nooijen}
\affiliation{Department of Chemistry, University of Waterloo, Waterloo, Ontario, N2L 3G1, Canada}

\begin{abstract}
	It was recently shown that path integral Monte Carlo can be used to directly compute partition functions of Hamiltonians with vibronic coupling [J.\@ Chem.\@ Phys.\@ \textbf{148}, 194110 (2018)].
	While the importance sampling Monte Carlo integration scheme was successful, it required many samples to reduce the stochastic error and suffered from the need to manually construct a sampling distribution for each system.
	We tackle these issues by using deterministic component selection for Gaussian mixture distributions (GMDs), introducing quasi-random numbers into the Monte Carlo sampling, and automatically optimizing the GMD parameters to obtain an improved sampling distribution.
	We demonstrate the effectiveness of our methods using vibronic model systems, but these methods are in principle widely applicable to general Monte Carlo sampling of GMDs.
\end{abstract}

\maketitle

\section{Introduction}
\label{sec:introduction}

Although the Born--Oppenheimer approximation (BOA) has proven itself to be an invaluable tool in computational chemistry, certain classes of systems cannot be accurately described under its restrictions.\cite{kouppel1984multimode}
Systems with coupling between vibrational and electronic degrees of freedom (vibronic coupling) have regions of nonadiabaticity, where adiabatic surfaces approach each other in energy, possibly crossing; this situation is often observed in photochemical reactions.\cite{robb1995conical,wodtke2004electronically,worth2004beyond,goel2016proposed}
Examples in biologically relevant molecules include the photoisomerization of retinal and the photostability of DNA base pairs.\cite{hahn2000quantum,sobolewski2005tautomeric}

When vibronic coupling is present, it is necessary to use methods that are not bound by the BOA.
Several approaches use a vibronic description of the form
\begin{align}
	\label{eq:H-full}
	\hat{H}^{a a'}
	&= \delta_{a a'} \sum_{j=1}^N \frac{\omega_j}{2} (\hat{p}_j^2 + \hat{q}_j^2)
		+ E^{(0) a a'}
	\notag \\ &\qquad
		+ \sum_{j=1}^N E^{(1) a a'}_j \hat{q}_j
		+ \sum_{j=1}^N \sum_{j'=1}^N E^{(2) a a'}_{j j'} \hat{q}_j \hat{q}_{j'}
	\notag \\ &\qquad
		+ \sum_{j=1}^N \sum_{j'=1}^N \sum_{j''=1}^N E^{(3) a a'}_{j j' j''} \hat{q}_j \hat{q}_{j'} \hat{q}_{j''}
		+ \cdots,
\end{align}
where $A$ diabatic surfaces (labelled by $a = 1, \ldots, A$) describing $N$ vibrational degrees of freedom are coupled via the off-diagonal ($a \ne a'$) elements of $\hat{H}$.\cite{worth2004beyond,worth2009multidimensional,goel2016proposed}
For convenience, all the quantities in Eq.~\eqref{eq:H-full} (i.e.\ the position and momentum coordinates, as well as the various parameters, and hence also the resulting Hamiltonian operator) are taken to be dimensionless, with $\hbar = 1$.

In recent years, multiple proposals have been put forward to find properties of such vibronic Hamiltonians using path integrals, and in particular path integral Monte Carlo (PIMC).\cite{ananth2013mapping,menzeleev2014kinetically,duke2017mean,liu2018path,raymond2018path}
While most of these focus primarily on dynamical properties, in Ref.~\onlinecite{raymond2018path} it was shown that PIMC may be used to compute the canonical partition function $Z$ of a system with vibronic coupling.

In the present work, we build on the approach established in Ref.~\onlinecite{raymond2018path}, offering several improvements.
To reduce the overall stochastic error, we augment the random sampling of a Gaussian mixture distribution (GMD) with two deterministic techniques.
The first is a way to select the component of the GMD for optimal reduction of the variance of the mean.
The other is quasi-Monte Carlo, which uses low-discrepancy sequences (quasi-random numbers) instead of pseudo-random numbers, and has seen success in physical, chemical, and financial applications.\cite{joy1996quasi,caflisch1998monte,chi2005scrambled,brown2013self,georgescu2013ground}

Furthermore, we describe a generic approach for optimizing the parameters of the GMD used in importance sampling to ensure that regions of high integrand magnitude can be sufficiently well explored, even when their locations are not already known.
This is accomplished by means of the simultaneous perturbation stochastic approximation (SPSA) algorithm, which repeatedly varies the GMD parameters in order to minimize a loss function.\cite{spall1992multivariate,spall1998implementation}

The remainder of this article is structured as follows: in Sec.~\ref{sec:background}, we provide the theoretical background for the subsequent sections; in Sec.~\ref{sec:methods}, we explain the proposed enhancements; in Sec.~\ref{sec:results}, we apply these enhancements to model systems; and in Sec.~\ref{sec:conclusions}, we give some concluding remarks.

\section{Background}
\label{sec:background}

To bring the reader up to speed with the variant of vibronic PIMC discussed in this work, we include a brief derivation in Sec.~\ref{sec:background-vibronic}.
Since randomized quasi-Monte Carlo is not frequently seen in computational chemistry literature, we give a short overview in Sec.~\ref{sec:background-rqmc}.
In Sec.~\ref{sec:background-spsa}, we review the simultaneous perturbation stochastic approximation algorithm.

\subsection{Vibronic path integral Monte Carlo}
\label{sec:background-vibronic}

The goal of Ref.~\onlinecite{raymond2018path} is to compute the partition function
\begin{align}
	Z
	&= \Tr e^{-\beta \hat{H}}
\end{align}
of the vibronic Hamiltonian $\hat{H}$ in Eq.~\eqref{eq:H-full} at reciprocal temperature $\beta = 1 / k_\mathrm{B} T$.
This is accomplished by inserting resolutions of the identity
\begin{align}
	\hat{\mathds{1}}
	&= \sum_{a=1}^A \int\! \dd{\vec{q}} \, \dyad{a \, \vec{q}}
\end{align}
in the combined diabatic basis and (normal mode) position representation to arrive at the expression
\begin{align}
	Z
	&= \sum_{a_1=1}^A \cdots \sum_{a_P=1}^A \int\! \dd{\vec{q}_1} \cdots \int\! \dd{\vec{q}_P}
		\mel{a_P \, \vec{q}_P}{e^{-\tau \hat{H}}}{a_1 \, \vec{q}_1}
	\notag \\ &\qquad\qquad\qquad\qquad\qquad
		\times \prod_{i=1}^{P-1} \mel{a_i \, \vec{q}_i}{e^{-\tau \hat{H}}}{a_{i+1} \, \vec{q}_{i+1}},
\end{align}
which has the form of a discretized imaginary time path integral with $P$ beads and time step $\tau = \beta / P$.
The Trotter factorization is then applied to obtain an approximation that is exact in the $P \to \infty$ limit.
It takes on the form
\begin{align}
	\label{eq:Z-g}
	Z
	&= \int\! \dd{\vec{R}} \, g(\vec{R}),
\end{align}
in which: we use $\vec{R}$ to mean the vector containing all the continuous path coordinates $\vec{q}_1$, $\ldots$, $\vec{q}_P$; the integrand is
\begin{align}
	\label{eq:gR}
	g(\vec{R})
	&= \Tr[ \mathds{O}(\vec{q}_P, \vec{q}_1) \mathds{M}(\vec{q}_1) \prod_{i=1}^{P-1} \mathds{O}(\vec{q}_i, \vec{q}_{i+1}) \mathds{M}(\vec{q}_{i+1}) ];
\end{align}
and the matrix-valued functions $\mathds{O}$ and $\mathds{M}$ have the elements
\begin{subequations}
\begin{align}
	\mathds{O}(\vec{q}, \vec{q}')_{a a'}
	&= \delta_{a a'} \mel{\vec{q}}{e^{-\tau \hat{h}^a}}{\vec{q}'}, \\
	\mathds{M}(\vec{q})_{a a'}
	&= \mel{a}{e^{-\tau \hat{V}(\vec{q})}}{a'}.
\end{align}
\end{subequations}
The operator $\hat{h}$ is diagonal in the diabatic basis and has the form
\begin{align}
	\label{eq:h}
	\hat{h}^a
	&= \sum_{j=1}^N \frac{\omega_j}{2} (\hat{p}_j^2 + \hat{q}_j^2)
		+ E^{(0) a a}
		+ \sum_{j=1}^N E^{(1) a a}_j \hat{q}_j,
\end{align}
whereas the operator
\begin{align}
	\hat{V}
	&= \hat{H} - \hat{h}
\end{align}
is diagonal in the position representation.
The shorthand notation
\begin{align}
	\hat{h}_\mathrm{o}
	&= \sum_{j=1}^N \frac{\omega_j}{2} (\hat{p}_j^2 + \hat{q}_j^2)
\end{align}
is sometimes used for the harmonic oscillator terms.

The integral in Eq.~\eqref{eq:Z-g} is then approximated by $N_\mathrm{MC}$ steps of Monte Carlo with importance sampling from the probability density function (pdf) $\pi(\vec{R})$:
\begin{align}
	Z
	&= \ev{f}_{\pi}
	= \int\! \dd{\vec{R}} \, \pi(\vec{R}) f(\vec{R})
	\approx \frac{1}{N_\mathrm{MC}} \sum_{i=1}^{N_\mathrm{MC}} f(\vec{R}_i)
	= \bar{f},
\end{align}
where
\begin{align}
	f(\vec{R})
	&= \frac{g(\vec{R})}{\pi(\vec{R})}.
\end{align}
In principle, $\pi(\vec{R})$ may be any normalized pdf that does not vanish on the support of $g(\vec{R})$.\footnote{
	The unnormalized function $\rho(\vec{R})$ that appears in Ref.~\onlinecite{raymond2018path} is related to $\pi(\vec{R})$ by the normalization: $\rho(\vec{R}) = Z_\rho \pi(\vec{R})$.
}
In practice, the distribution $\pi$ must be chosen to allow efficient sampling and to have significant overlap with $g$.
If the latter condition is not fulfilled, the result is spectacular failure of the method, as shown in Ref.~\onlinecite{raymond2018path}.

The ease and efficiency with which one can sample from a Gaussian mixture distribution (GMD), combined with the GMD-like form obtained for $g$ when $\hat{V} = 0$, make a GMD a natural choice for $\pi$.
The general form of a GMD pdf is the convex combination
\begin{align}
	\label{eq:mixture-distribution}
	\pi(\vec{R})
	&= \sum_{b=1}^B w^b \pi^b(\vec{R})
\end{align}
of multivariate Gaussian pdfs $\pi^b(\vec{R})$.
The straightforward approach for randomly sampling a point $\vec{R}$ from $\pi$ is to first choose a component $\pi^b$, and then sample from $\pi^b$ according to its mean vector $\vec{d}^b$ and covariance matrix $\vec{\Sigma}^b$.
Although each component pdf
\begin{align}
	\label{eq:gaussian-component}
	\pi^b(\vec{R})
	&= \frac{1}{\sqrt{(2\pi)^{P N} \det \vec{\Sigma}^b}} e^{-\frac{1}{2} (\vec{R} - \vec{d}^b)^\mathrm{T} (\vec{\Sigma}^b)^{-1} (\vec{R} - \vec{d}^b)}
\end{align}
has a fixed number of parameters (the mean vector $\vec{d}^b$ has length $P N$ and the covariance matrix $\vec{\Sigma}^b$ is $P N \times P N$), the total number of parameters is arbitrary, scaling linearly with $B$.

Broadly speaking, there are three issues to be addressed when using the GMD $\pi$ as a sampling distribution: how to select a component $\pi^b$, how to sample from $\pi^b$, and how to choose the parameters of each component.
In this work, we consider some improvements to all of these areas.

\subsection{Randomized quasi-Monte Carlo}
\label{sec:background-rqmc}

It is often more important for Monte Carlo (MC) that sampled points be consistently distributed rather than randomly distributed.
To take advantage of this, low-discrepancy sequences (LDSs) may be used as a smoother substitute for pseudo-random sequences, such as those output by a pseudo-random number generator (RNG).
The aim of an LDS is to produce values which are as evenly spaced as possible within some volume (typically the $D$-dimensional hypercube $[0, 1)^D$).\cite{niederreiter1988low}
In this context, the values forming an LDS are called ``quasi-random numbers'' and their application to MC gives rise to quasi-Monte Carlo (qMC\footnote{
	We use a lower-case ``q'' in ``qMC'' to avoid confusion with quantum Monte Carlo, which is frequently abbreviated as ``QMC''.
}).\cite{niederreiter1978quasi}
The chief argument in favor of qMC is that the estimates it provides have an error whose asymptotic scaling with the number of samples $N_\mathrm{MC}$ is expected to be better than the $N_\mathrm{MC}^{-\frac{1}{2}}$ scaling seen in plain MC.\cite{birge1995quasi,caflisch1998monte}

In the present work, we employ Sobol sequences to produce quasi-random numbers.\cite{sobol1967distribution}
Other sequences may also be used, such as those of Halton or Faure, but Sobol sequences have been observed to perform better for high-dimensional problems\cite{morokoff1995quasi} and there exists a convenient software package to generate them.\cite{soboljl}
Some authors suggest skipping the initial points of a low-discrepancy sequence,\cite{acworth1998comparison} but we have noticed no change when doing so, and therefore choose not to skip any points.
We have included a demonstration of qMC in Appendix~\ref{sec:qmc-demo}, where we show its application to the classic MC problem of estimating the number $\pi$, and a very brief discussion in Appendix~\ref{sec:qmc-gaussian}, where we mention some associated difficulties with Gaussian distributions and correlations in many dimensions.
For more information about qMC, the reader is directed to Refs.~\onlinecite{niederreiter1978quasi,caflisch1998monte,owen2003quasi}.

The primary concern with qMC is its inability to provide error estimates.
Some theoretical bounds are known, but they are not useful in practice.\cite{tuffin2004randomization}
Instead, one may use randomized quasi-Monte Carlo (RqMC), which reintroduces pseudo-random numbers in order to compute statistical error bars.
The idea is simple: several low-discrepancy sequences are run in tandem, each with a different shift $\vec{v}$ generated by an RNG.\cite{tuffin2004randomization,okten2004randomized}
An LDS where every point $\vec{u}$ is shifted by the same displacement vector $\vec{v}$ to produce
\begin{align}
	\vec{w}'
	&= \vec{u} + \vec{v}
\end{align}
is still an LDS in the new shifted hypercube $[0, 1)^D + \vec{v}$.
If we take the elements of $\vec{w}'$ that stick out of the original hypercube $[0, 1)^D$ and wrap them back in as if using periodic boundary conditions, we end up with the vector $\vec{w}$, whose elements (for $i = 1, \ldots, D$) are
\begin{align}
	w_i
	&= w'_i \rem 1
	= w'_i - \lfloor w'_i \rfloor.
\end{align}
Here, $x \rem y$ is the unique value in $[0, y)$ such that $x - (x \rem y)$ is an integer multiple of $y > 0$ (commonly referred to as the remainder), and $\lfloor x \rfloor$ is the largest integer not exceeding $x$ (commonly called the floor).

After being shifted and wrapped, every individual sequence in RqMC should still be evenly distributed in the original hypercube, and will generate a single estimate for the quantity in question.
Because the estimates depend on the random shifts, they are themselves random variables and may be combined in the customary ways to obtain not only a sample mean, but also its standard error.
This adds another parameter into the calculation: the number of shifted sequences $N_\mathrm{S}$ must be chosen carefully to strike a balance between ensuring a sufficient sample size for valid estimation of the error, and conserving the smooth results provided by low-discrepancy sequences.
An example of this trade-off is shown in Appendix~\ref{sec:rqmc}.

In addition to drawing uniform samples from a hypercube, we require the ability to sample quasi-randomly from the multivariate Gaussian pdf in Eq.~\eqref{eq:gaussian-component}.
To do this, we first generate a point in the $D$-dimensional hypercube $[0, 1)^D$ using a $D$-dimensional Sobol sequence and shift this point to randomize it.
Then each of the $D$ elements is treated as a cumulative distribution function (cdf) value (i.e.\ as a probability) and inverted to produce a one-dimensional standard Gaussian sample.\footnote{
	Some programming languages and libraries include a function called \textsc{erfcinv}, which computes the inverse of the complementary error function and may be used for Gaussian cdf inversion: $\Phi^{-1}(\phi) = -\sqrt{2} \erfcinv(2 \phi)$.
}
Finally, these are combined to generate the appropriate multi-dimensional sample.
In full, the randomized low-discrepancy multivariate Gaussian sampling routine \textsc{sample\_q}($b$) may be implemented as in Alg.~\ref{alg:sampleq}, where $b = 1, \ldots, B$ determines the distribution $\pi^b$.

\begin{figure}
\begin{algorithm}[H]
	\caption{Multivariate Gaussian sampling for randomized quasi-Monte Carlo.}
	\label{alg:sampleq}

	\begin{algorithmic}[0]
		\Require $N_\mathrm{S} \ge 1$
		\Require mean vector $\vec{d}^b$
		\Require covariance matrix $\vec{\Sigma}^b$ whose inverse $(\vec{\Sigma}^b)^{-1}$ has eigenvalues $\vec{\Lambda}^b$ and eigenvectors $\vec{S}^b$
		\Require shift vectors $\vec{v}^b_s$
		\Require \Call{next}{$b$} returns the next point in the $D$-dimensional LDS labelled by $b$
		\Require \Call{cdfinv}{$w$} performs Gaussian cdf inversion
		\Statex
		\Function{sample\_q}{$b$}
			\State $\vec{u} \gets$ \Call{next}{$b$}
			\Statex
			\For{$s \gets 1:N_\mathrm{S}$}
				\For{$i \gets 1:D$}
					\State $w_{s i} \gets (u_i + v^b_{s i}) \rem 1$
					\State $y_{s i} \gets$ \Call{cdfinv}{$w_{s i}$} $/ \sqrt{\Lambda^b_i}$
				\EndFor
				\State $\vec{x}_s \gets \vec{S}^b \vec{y}_s + \vec{d}^b$
			\EndFor
			\Statex
			\State \Call{return}{$\vec{x}$}
		\EndFunction
	\end{algorithmic}
\end{algorithm}
\end{figure}

\subsection{Simultaneous perturbation stochastic approximation}
\label{sec:background-spsa}

The goal of the simultaneous perturbation stochastic approximation (SPSA) algorithm is to find the global minimum of a loss function $\mathcal{L}(\vec{\Theta})$ by varying the $G$-dimensional parameter vector $\vec{\Theta}$.\cite{spall1992multivariate}
Although similar to the finite difference stochastic approximation (FDSA) algorithm, in which each dimension requires a finite difference evaluation at every step, SPSA perturbs all $G$ dimensions simultaneously, accelerating convergence.
The ``SA'' part of the name refers to the approximation made in computing the gradient of $\mathcal{L}$ using samples of
\begin{align}
	\ell(\vec{\Theta})
	&= \mathcal{L}(\vec{\Theta}) + \mathrm{noise},
\end{align}
in contrast to many other methods that require exact gradients.\cite{spall1998implementation}

This iterative algorithm is straightforward to describe and implement.
The function \textsc{spsa}(\textsc{loss}, $k$, $\vec{\Theta}$) in Alg.~\ref{alg:spsa} is repeatedly called with increasing integer $k \ge 1$ until convergence or a maximum iteration threshold $N_\mathrm{SPSA}$.
The argument \textsc{loss} is a function that provides an estimate of the loss, together with the standard error of the estimate (which is not used here, but will be necessary elsewhere).
Implicit in the presented algorithm is our choice to use a symmetric Bernoulli distribution to generate the perturbation vector $\vec{\Delta}$.

\begin{figure}
\begin{algorithm}[H]
	\caption{One iteration of SPSA.}
	\label{alg:spsa}

	\begin{algorithmic}[0]
		\Require \Call{flip}{{}} returns $\pm 1$, each with probability $1/2$
		\Statex
		\Function{spsa}{\textsc{loss}, $k$, $\vec{\Theta}$}
			\For{$i \gets 1:G$}
				\Comment{Perturb parameters.}

				\State $\Delta_i \gets$ \Call{flip}{{}}
				\State $\Theta^+_i \gets \Theta_i + c_k \Delta_i$
				\State $\Theta^-_i \gets \Theta_i - c_k \Delta_i$
			\EndFor
			\Statex
			\State $\ell^+, \sigma^+ \gets$ \Call{loss}{$\vec{\Theta}^+$}
			\Comment{Approximate gradient.}
			\State $\ell^-, \sigma^- \gets$ \Call{loss}{$\vec{\Theta}^-$}
			\State $\vec{g} \gets \frac{\ell^+ - \ell^-}{2 c_k} \vec{\Delta}$
			\Statex
			\State $\vec{\Theta}' \gets \vec{\Theta} - a_k \vec{g}$
			\Comment{Update parameters.}
			\State \Call{return}{$\vec{\Theta}'$}
		\EndFunction
	\end{algorithmic}
\end{algorithm}
\end{figure}

Besides $N_\mathrm{SPSA}$, the algorithm requires five parameters to be chosen: $A$, $a$, $c$, $\alpha$, and $\gamma$.
These are used to compute the gain sequences
\begin{subequations}
\label{eq:spsa-gain}
\begin{align}
	a_k
	&= \frac{a}{(A+k)^\alpha} \\
\intertext{and}
	c_k
	&= \frac{c}{k^\gamma}.
\end{align}
\end{subequations}
For $\alpha$ and $\gamma$, we use the recommended values $\alpha = 0.602$ and $\gamma = 0.101$, and we set $A$ to \SI{10}{\percent} of $N_\mathrm{SPSA}$ and $c$ to be equal to the standard error of $\ell(\vec{\Theta})$ before the initial iteration.\cite{spall1998implementation}
For $a$, we resort to manual adjustment; the optimization will be inefficient if it is too small, and erratic if it is too large.

\section{Methods}
\label{sec:methods}

In this section, we present our proposed improvements to GMD sampling for purposes of estimating partition functions of vibronic Hamiltonians using PIMC.
In Sec.~\ref{sec:methods-component} and Sec.~\ref{sec:methods-quasi}, we describe how to make the sampling more efficient.
In Sec.~\ref{sec:methods-optimization} and Sec.~\ref{sec:methods-optimization-deformation}, we show how to optimize the GMD parameters to better describe the integrand.

\subsection{Deterministic component selection}
\label{sec:methods-component}

The intuitive way to sample from a mixture distribution of the form in Eq.~\eqref{eq:mixture-distribution} is to first choose a component $\pi^b$ using an RNG and then sample from that component.
Repeating this $N_\mathrm{MC}$ times results in the point set $\{ \vec{R}_i \}_{i=1}^{N_\mathrm{MC}}$, where it is assumed that the points are not correlated.
If the components are chosen in accordance with the weights $w^b$, these points can be used in the usual way\cite{young2015everything} to find the sample mean
\begin{align}
	\label{eq:simple-mean}
	\bar{f}
	&= \frac{1}{N_\mathrm{MC}} \sum_{i=1}^{N_\mathrm{MC}} f(\vec{R}_i),
\end{align}
sample variance
\begin{align}
	\label{eq:simple-var}
	s_f^2
	&= \frac{1}{N_\mathrm{MC}-1} \sum_{i=1}^{N_\mathrm{MC}} (f(\vec{R}_i) - \bar{f})^2,
\end{align}
and sample error of the mean (also known as standard error)
\begin{align}
	\label{eq:simple-error}
	s_{\bar{f}}
	&= \sqrt{\frac{s_f^2}{N_\mathrm{MC}}}
	= \sqrt{\frac{1}{N_\mathrm{MC} (N_\mathrm{MC}-1)} \sum_{i=1}^{N_\mathrm{MC}} (f(\vec{R}_i) - \bar{f})^2}.
\end{align}
Although quite straightforward (see Alg.~\ref{alg:basic-sampling} and Fig.~\ref{fig:example-basic-sampling}), this approach struggles with small component weights.
For example, consider $N_\mathrm{MC} = 10^6$ and $w^{b^\star} = 10^{-6}$ for some $b^\star$.
From a simple binomial distribution analysis, we expect to find exactly one sample for this component only
\begin{align}
	\binom{10^6}{1} (10^{-6})^1 (1 - 10^{-6})^{10^6 - 1}
	&\approx \SI{37}{\percent}
\end{align}
of the time, with it being under-represented (zero samples) \SI{37}{\percent} of the time, and over-represented (two or more samples) the remaining \SI{26}{\percent}.\footnote{
	The asymptotic distribution as $N_\mathrm{MC} \to \infty$ and $w^b N_\mathrm{MC} = 1$ is a Poisson distribution with mean 1, resulting in the following probabilities: $1/e$ for zero samples, $1/e$ for one sample, and $1-2/e$ for two or more samples.
}
In any event, it is unlikely that just a handful of samples is sufficient to properly explore the component.

\begin{figure}
\begin{algorithm}[H]
	\caption{GMD sampling with stochastic component selection.}
	\label{alg:basic-sampling}

	\begin{algorithmic}[0]
		\Require $N_\mathrm{MC} \ge 1$
		\Require \Call{choose}{$\vec{w}$} returns index $b$ with probability $w^b$
		\Require \Call{sample}{$b$} returns a sample from $\pi^b$
		\Statex
		\For{$i \gets 1:N_\mathrm{MC}$}
			\State $b \gets$ \Call{choose}{$\vec{w}$}
			\Comment{Component selection.}
			\Statex
			\State $\vec{R}_i \gets$ \Call{sample}{$b$}
			\Comment{Component sampling.}
			\State $f_i \gets f(\vec{R}_i)$
		\EndFor
	\end{algorithmic}
\end{algorithm}
\end{figure}

\begin{figure}
	\includegraphics{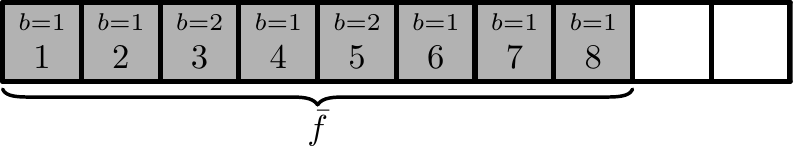}
	\caption{
		Example of $N_\mathrm{MC} = 8$ samples $f_i$ drawn using Alg.~\ref{alg:basic-sampling}, with GMD components stochastically chosen in the order 1, 1, 2, 1, 2, 1, 1, 1.
		The numbers along the bottom indicate the order in which the samples were obtained, and the $b$ labels show the GMD components from which they were sampled.
		The brace shows the values used to compute the average $\bar{f}$.
	}
	\label{fig:example-basic-sampling}
\end{figure}

The formulas in Eqs.~\eqref{eq:simple-mean}--\eqref{eq:simple-error} continue to hold so long as the effective distribution of components is consistent with the weights, even if the components are chosen in some other way, such as by a deterministic procedure.
This suggests that we could sample from each component exactly $N^b = w^b N_\mathrm{MC}$ times, in any order.
The obvious flaw with this approach is that $N^b$ must be an integer, so a very small $w^b$ implies a very large $N_\mathrm{MC}$, which is inconvenient.
In the remainder of this section, we demonstrate how to adjust the standard formulas to compensate for point sets where the distribution of components used for sampling does not match their weights, and provide a strategy for choosing the components during sampling.

Suppose we sample from each component $\pi^b$ an arbitrary number of times $N^b$, obtaining the point set $\bigcup_{b=1}^B \{ \vec{R}^b_i \}_{i=1}^{N^b}$.
From these samples, we may compute a collection of sample averages
\begin{align}
	\bar{f}^b
	&= \frac{1}{N^b} \sum_{i=1}^{N^b} f(\vec{R}^b_i).
\end{align}
The combined quantity
\begin{align}
	\label{eq:comb-mean}
	\bar{f}
	&= \sum_{b=1}^B w^b \bar{f}^b
\end{align}
is an unbiased estimator for $\ev{f}_\pi$:
\begin{align}
	\ev{\bar{f}}
	&= \sum_{b=1}^B w^b \frac{1}{N^b} \sum_{i=1}^{N^b} \ev{f}_{\pi^b}
	= \sum_{b=1}^B w^b \int\! \dd{\vec{R}} \, \pi^b(\vec{R}) f(\vec{R})
	\notag \\ &
	= \int\! \dd{\vec{R}} \, \pi(\vec{R}) f(\vec{R})
	= \ev{f}_\pi.
\end{align}

Each $\bar{f}^b$ is itself an independent random variable with population variance $\sigma_{\bar{f}^b}^2$.
As shown in Appendix~\ref{sec:variance-estimator}, the overall population variance of $\bar{f}$ is then given by
\begin{align}
	\sigma_{\bar{f}}^2
	&= \sum_{b=1}^B (w^b)^2 \sigma_{\bar{f}^b}^2.
\end{align}
For the variance $\sigma^2_{\bar{f}^b}$, we have the usual unbiased estimate
\begin{align}
	s_{\bar{f}^b}^2
	&= \frac{s_{f^b}^2}{N^b}
	= \frac{1}{N^b (N^b-1)} \sum_{i=1}^{N^b} (f(\vec{R}^b_i) - \bar{f}^b)^2,
\end{align}
from which we get
\begin{align}
	\label{eq:comb-var}
	s_{\bar{f}}^2
	&= \sum_{b=1}^B (w^b)^2 s_{\bar{f}^b}^2.
\end{align}
Thus, the standard error may be estimated by
\begin{align}
	s_{\bar{f}}
	&= \sqrt{\sum_{b=1}^B (w^b)^2 s_{\bar{f}^b}^2}
	= \sqrt{\sum_{b=1}^B \frac{(w^b)^2 s_{f^b}^2}{N^b}}.
\end{align}

Although it is possible to fix $N^b$ ahead of time, it is not necessary to do so.
Instead, we may choose each component during sampling in a way that attempts to optimally reduce the overall statistical error.
We observe that after sampling $N^b$ points from each component, an additional sample from the component labelled by $b^\star$ changes the sample variance of the mean $s_{\bar{f}}^2$ to
\begin{align}
	(s_{\bar{f}}')^2
	&= \frac{(w^{b^\star})^2 (s_{f^{b^\star}}')^2}{N^{b^\star} + 1}
		+ \smashoperator{\sum_{\substack{b=1 \\ (b \ne b^\star)}}^B} \frac{(w^b)^2 s_{f^b}^2}{N^b}.
\end{align}
Ideally, the estimate of the component variance will not be greatly affected by a single sample, so we make the simplifying assumption that
\begin{align}
	(s_{f^{b^\star}}')^2
	&\approx s_{f^{b^\star}}^2.
\end{align}
This allows us to write
\begin{align}
	(s_{\bar{f}}')^2
	&\approx s_{\bar{f}}^2
		- (w^{b^\star})^2 s_{f^{b^\star}}^2 \left[ \frac{1}{N^{b^\star}} - \frac{1}{N^{b^\star} + 1} \right],
\end{align}
from which it follows that
\begin{align}
	\label{eq:var-change}
	(s_{\bar{f}}')^2
	&\approx s_{\bar{f}}^2
		- \frac{(w^{b^\star})^2 s_{\bar{f}^{b^\star}}^2}{N^{b^\star} + 1},
\end{align}
and we see that to lower the standard error most quickly, it is beneficial to choose the component with the largest
\begin{align}
	\label{eq:component-selection}
	\delta^b
	&= \frac{(w^b)^2 s_{\bar{f}^b}^2}{N^b + 1}.
\end{align}
This criterion makes intuitive sense, as it targets components that have larger weights (because they are more important to sample well), larger errors of the mean (because they have a lot of remaining uncertainty), and fewer samples (because they have not been explored as thoroughly).
Since the computation of $\delta^b$ requires an existing estimate of the variance in that component, it is necessary to bootstrap this method by taking $N_\mathrm{boot}$ points from each component.

When implementing the above prescription, one must be careful to avoid the common formula in Eq.~\eqref{eq:simple-var} for estimating variance, since it scales linearly with the number of samples.
Using it after sampling each point leads to quadratic scaling of the total sampling algorithm.
To avoid this undesirable behavior, one may use an efficient ``online'' variance update scheme.
For example, the scheme of Youngs and Cramer for the $L$-point sample variance $s_f^2(L)$ is
\begin{subequations}
\label{eq:var-update}
\begin{align}
	s_f^2(L)
	&= \frac{U_f(L)}{L-1},
\end{align}
where
\begin{align}
	T_f(L)
	&= T_f(L-1) + f(\vec{R}_L), \\
	U_f(L)
	&= U_f(L-1) + \frac{(L f(\vec{R}_L) - T_f(L))^2}{L (L-1)},
\end{align}
\end{subequations}
subject to $T_f(0) = U_f(1) = 0$.\cite{youngs1971some,chan1983algorithms}
These updates can be evaluated in constant time at each step of the calculation.
The scheme is made more effective with a conditioning step where the data are shifted by the mean:\cite{chan1983algorithms}
\begin{align}
	f(\vec{R}_L)
	&\to \Delta f_L = f(\vec{R}_L) - \bar{f}.
\end{align}
Although the final sample mean will not be known until the end of the calculation, uniformly shifting the data has no impact on its variance.
Thus, the precise value of the shift is not crucial, and the estimates $\bar{f}_\mathrm{boot}^b$ obtained from the bootstrap points may be used.
The overall algorithm for deterministic component selection is shown in Alg.~\ref{alg:component}, with an example in Fig.~\ref{fig:example-component}.

\begin{figure}
\begin{algorithm}[H]
	\caption{GMD sampling with deterministic component selection.}
	\label{alg:component}

	\begin{algorithmic}[0]
		\Require $N_\mathrm{boot} \ge 2$
		\Require $N_\mathrm{MC} \ge B N_\mathrm{boot}$
		\Require \Call{sample}{$b$} returns a sample from $\pi^b$
		\Statex
		\For{$b \gets 1:B$}
			\Comment{Bootstrap.}
			\For{$i \gets 1:N_\mathrm{boot}$}
				\State $\vec{R}^b_i \gets$ \Call{sample}{$b$}
				\State $f^b_i \gets f(\vec{R}^b_i)$
			\EndFor
			\Statex
			\State $N^b \gets N_\mathrm{boot}$
			\State $\bar{f}^b_\mathrm{boot} \gets \frac{1}{N^b} \sum_{i=1}^{N^b} f^b_i$
			\Statex
			\State $T^b \gets 0$
			\State $U^b \gets \sum_{i=1}^{N^b} (f^b_i - \bar{f}^b_\mathrm{boot})^2$
			\State $s_{\bar{f}^b}^2 \gets \frac{U^b}{N^b (N^b-1)}$
		\EndFor
		\Statex
		\For{$i \gets 1:(N_\mathrm{MC} - B N_\mathrm{boot})$}
			\Comment{Sampling.}
			\For{$b \gets 1:B$}
				\Comment{Component selection.}
				\State $\delta^b \gets \frac{(w^b)^2 s_{\bar{f}^b}^2}{N^b + 1}$
			\EndFor
			\State $b \gets$ \Call{argmax}{$\vec{\delta}$}
			\State $N^b \gets N^b + 1$
			\Statex
			\State $\vec{R}^b_{N^b} \gets$ \Call{sample}{$b$}
			\Comment{Component sampling.}
			\State $f^b_{N^b} \gets f(\vec{R}^b_{N^b})$
			\Statex
			\State $\Delta f \gets f^b_{N^b} - \bar{f}^b_\mathrm{boot}$
			\Comment{Online variance update.}
			\State $T^b \gets T^b + \Delta f$
			\State $U^b \gets U^b + \frac{(N^b \Delta f - T^b)^2}{N^b (N^b-1)}$
			\State $s_{\bar{f}^b}^2 \gets \frac{U^b}{N^b (N^b-1)}$
		\EndFor
	\end{algorithmic}
\end{algorithm}
\end{figure}

\begin{figure}
	\includegraphics{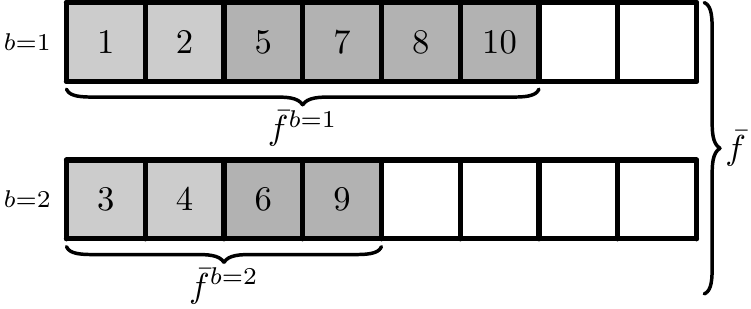}
	\caption{
		Example of $N_\mathrm{MC} = 10$ samples $f^b_i$ drawn using Alg.~\ref{alg:component}, with GMD components deterministically chosen in the order 1, 1, 2, 2 for the bootstrap ($N_\mathrm{boot} = 2$), followed by 1, 2, 1, 1, 2, 1.
		The component-wise counts are $N^{b=1} = 6$ and $N^{b=2} = 4$.
		The numbers indicate the order in which the samples were obtained, and the $b$ labels on the left identify the GMD components.
		The braces show the way in which the samples are combined: those for each GMD component are first grouped into $\bar{f}^b$, and these averages are subsequently used to form $\bar{f}$.
	}
	\label{fig:example-component}
\end{figure}

\subsection{Randomized quasi-Monte Carlo}
\label{sec:methods-quasi}

\begin{figure}
	\includegraphics{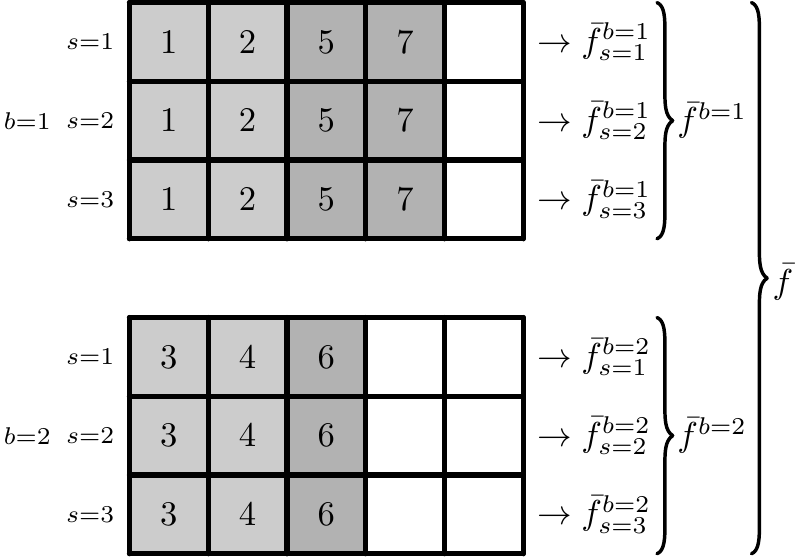}
	\caption{
		Example of $N_\mathrm{MC} = 21$ samples $f^b_{i s}$ across $N_\mathrm{S} = 3$ sequences drawn using Alg.~\ref{alg:rqmc} from $N_\mathrm{qMC} = 7$ low-discrepancy points, with GMD components deterministically chosen in the order 1, 1, 2, 2 for the bootstrap ($N_\mathrm{boot} = 2$), followed by 1, 2, 1.
		The component-wise counts are $N^{b=1} = 4$ and $N^{b=2} = 3$.
		The numbers indicate the order in which the samples were obtained, and the $b$ and $s$ labels on the left identify the GMD components and randomized sequences.
		The arrows and braces show the way in which the samples are combined: each sequence is first folded into the pseudo-random quantity $\bar{f}^b_s$, then these are averaged for each GMD component to find $\bar{f}^b$, and finally the GMD component averages are combined into $\bar{f}$.
	}
	\label{fig:example-rqmc}
\end{figure}

In order to incorporate quasi-random numbers into the PIMC study of vibronic Hamiltonians, the essential change is the substitution of an RNG by $N_\mathrm{S}$ randomly shifted Sobol sequences for each GMD component.
Instead of drawing $N_\mathrm{MC}$ pseudo-random points from an RNG, one then obtains $N_\mathrm{qMC}$ quasi-random ones from the low-discrepancy sequences.
The main consequence of this change can be seen by comparing the examples in Fig.~\ref{fig:example-component} and Fig.~\ref{fig:example-rqmc}: the component averages $\bar{f}^b$ are not computed from $N^b$ values, but from $N_\mathrm{S}$ of them.
This does not affect the prior formulas in Eq.~\eqref{eq:comb-mean} and Eq.~\eqref{eq:comb-var} for combining these averages into the overall mean $\bar{f}$ and variance $s^2_{\bar{f}}$.

However, the component selection criterion in Eq.~\eqref{eq:component-selection} and the online variance update formula in Eq.~\eqref{eq:var-update} are not directly applicable when RqMC is in use.
They are based on the premise that each additional sample increases the number of entities being considered.
Since the number of statistically independent values used for computing each component mean $\bar{f}^b$ is now always fixed to be $N_\mathrm{S}$, the reasoning behind these equations is no longer valid.
Conveniently, the online variance update is not necessary anymore, as the conventional variance formula becomes independent of the number of samples and can therefore be computed in constant time.

Component selection, on the other hand, becomes highly non-trivial, since Eq.~\eqref{eq:var-change} changes to
\begin{align}
	(s_{\bar{f}}')^2
	&= s_{\bar{f}}^2
		- \frac{(w^{b^\star})^2}{N_\mathrm{S}} \left( s_{f^{b^\star}}^2 - (s_{f^{b^\star}}')^2 \right),
\end{align}
which requires us to be able to predict the change in variance for a component when a single quasi-random sample is added.
To side-step this issue altogether, we argue that the intuition behind Eq.~\eqref{eq:component-selection} is still legitimate and we continue to choose the component with the largest
\begin{align}
	\delta^b
	&= \frac{(w^b)^2 s_{\bar{f}^b}^2}{N^b + 1},
\end{align}
where $N^b$ is the total number of quasi-random points drawn so far for that component.
The overall procedure for RqMC with deterministic component selection is detailed in Alg.~\ref{alg:rqmc}, with an example in Fig.~\ref{fig:example-rqmc}.
Note that $N_\mathrm{qMC}$ is the number of quasi-random points obtained in total from the LDSs; for a fair comparison with plain Monte Carlo methods, we also define $N_\mathrm{MC} = N_\mathrm{S} N_\mathrm{qMC}$, which is the total number of times that $f(\vec{R})$ is evaluated.

\begin{figure}
\begin{algorithm}[H]
	\caption{Quasi-random GMD sampling.}
	\label{alg:rqmc}

	\begin{algorithmic}[0]
		\Require $N_\mathrm{boot} \ge 1$
		\Require $N_\mathrm{S} \ge 2$
		\Require $N_\mathrm{qMC} \ge B N_\mathrm{boot}$
		\Require \Call{sample\_q}{$b$} is implemented as in Alg.~\ref{alg:sampleq} in Sec.~\ref{sec:background-rqmc}
		\Statex
		\For{$b \gets 1:B$}
			\Comment{Bootstrap.}
			\For{$i \gets 1:N_\mathrm{boot}$}
				\State $\vec{R}^b_i \gets$ \Call{sample\_q}{$b$}
				\For{$s \gets 1:N_\mathrm{S}$}
					\State $f^b_{i s} \gets f(\vec{R}^b_{i s})$
				\EndFor
			\EndFor
			\State $N^b \gets N_\mathrm{boot}$
			\Statex
			\For{$s \gets 1:N_\mathrm{S}$}
				\State $T_s^b \gets \sum_{i=1}^{N^b} f^b_{i s}$
				\State $\bar{f}^b_s \gets \frac{T_s^b}{N^b}$
			\EndFor
			\Statex
			\State $\bar{f}^b \gets \frac{1}{N_\mathrm{S}} \sum_{s=1}^{N_\mathrm{S}} \bar{f}^b_s$
			\State $s_{\bar{f}^b}^2 \gets \frac{1}{N_\mathrm{S} (N_\mathrm{S}-1)} \sum_{s=1}^{N_\mathrm{S}} (\bar{f}^b_s - \bar{f}^b)^2$
		\EndFor
		\Statex
		\For{$i \gets 1:(N_\mathrm{qMC} - B N_\mathrm{boot})$}
			\Comment{Sampling.}
			\For{$b \gets 1:B$}
				\Comment{Component selection.}
				\State $\delta^b \gets \frac{(w^b)^2 s_{\bar{f}^b}^2}{N^b + 1}$
			\EndFor
			\State $b \gets$ \Call{argmax}{$\vec{\delta}$}
			\State $N^b \gets N^b + 1$
			\Statex
			\State $\vec{R}^b_{N^b} \gets$ \Call{sample\_q}{$b$}
			\Comment{Component sampling.}
			\For{$s \gets 1:N_\mathrm{S}$}
				\State $f^b_{N^b s} \gets f(\vec{R}^b_{N^b s})$
				\State $T_s^b \gets T_s^b + f^b_{N^b s}$
				\State $\bar{f}^b_s \gets \frac{T_s^b}{N^b}$
			\EndFor
			\Statex
			\State $\bar{f}^b \gets \frac{1}{N_\mathrm{S}} \sum_{s=1}^{N_\mathrm{S}} \bar{f}^b_s$
			\Comment{Variance update.}
			\State $s_{\bar{f}^b}^2 \gets \frac{1}{N_\mathrm{S} (N_\mathrm{S}-1)} \sum_{s=1}^{N_\mathrm{S}} (\bar{f}^b_s - \bar{f}^b)^2$
		\EndFor
	\end{algorithmic}
\end{algorithm}
\end{figure}

\clearpage

\subsection{Parameter optimization}
\label{sec:methods-optimization}

A major concern with the sampling scheme presented in Ref.~\onlinecite{raymond2018path} is the need to choose a sampling GMD $\pi$ that closely approximates the true path density $g$.
Neglecting to do so results in catastrophic failure when estimating $\ev{f}_\pi$, as the sampled points are chosen to lie in irrelevant locations that contribute nothing of substance to the integral in Eq.~\eqref{eq:Z-g}.
Thus, we need a way to refine $\pi$, making it more similar to $g$, and this requires a quantitative measurement of the difference between them.

A common metric to determine the similarity between two distributions is the relative entropy, also known as the Kullback--Leibler divergence.\cite{kullback1951information}
It is not symmetric in the distributions, so given pdfs $g(\vec{R})/Z$ and $\pi(\vec{R})$, there are two possible quantities:
\begin{subequations}
\begin{align}
	D(\pi || g/Z)
	&= \int\! \dd{\vec{R}} \, \pi(\vec{R}) \log \frac{\pi(\vec{R}) Z}{g(\vec{R})}
	\notag \\ &
	= \log \ev{f}_\pi - \ev{\log f}_\pi \\
\intertext{and}
	D(g/Z || \pi)
	&= \int\! \dd{\vec{R}} \, \frac{g(\vec{R})}{Z} \log \frac{g(\vec{R})}{\pi(\vec{R}) Z}
	\notag \\ &
	= \log \ev{\frac{1}{f}}_{g/Z} - \ev{\log \frac{1}{f}}_{g/Z}.
\end{align}
\end{subequations}
For our purposes, these expressions are both ill-defined, because $g(\vec{R})$ may differ in sign from $Z$ (see Appendix~\ref{sec:negative-g} for an example), and $g(\vec{R})/Z$ is therefore not necessarily a pdf.
This may be remedied by using
\begin{align}
	\frac{\abs{g(\vec{R})}}{Z_{||}}
	&= \frac{\abs{g(\vec{R})}}{\int\! \dd{\vec{R}'} \abs{g(\vec{R}')}}
\end{align}
instead; if the negative regions of $g$ are small in magnitude, this will be a good approximation to $g/Z$.
However, with no way to sample directly from $\abs{g}/Z_{||}$, it is not clear how to efficiently evaluate $D(\abs{g}/Z_{||} \,||\, \pi)$.

Thus, the best measure of quality that we can get is
\begin{align}
	\label{eq:relative-entropy}
	D(\pi \,||\, \abs{g}/Z_{||})
	&= \int\! \dd{\vec{R}} \, \pi(\vec{R}) \log \frac{\pi(\vec{R}) Z_{||}}{\abs{g(\vec{R})}}
	\notag \\ &
	= \log \ev{\abs{f}}_\pi - \ev{\log \abs{f}}_\pi.
\end{align}
We can then minimize this quantity by modifying $\pi$ using any of a number of optimization techniques, yielding
\begin{align}
	\argmin_{\vec{E}} D(\pi_{\vec{E}} \,||\, \abs{g}/Z_{||}),
\end{align}
where the sampling distribution $\pi_{\vec{E}}$ is parameterized by the tensor $\vec{E}$.
In most cases, attempting this would not be fruitful: if first we need to optimize $\pi$ in order to be able to compute $\ev{f}_\pi$, then we can't use Eq.~\eqref{eq:relative-entropy} directly.
However, for cases where $\pi$ is already a reasonably good approximation of $g/Z$, this could be used to improve it further.

Though each $\pi_{\vec{E}}$ may be any GMD, as specified by its weights, mean vectors, and covariance matrices, we choose to construct it from the diagonal Hamiltonian $\hat{h}_{\vec{E}}$ with the elements
\begin{align}
	\label{eq:gmd-hamiltonian}
	\hat{h}_{\vec{E}}^b
	&= \sum_{j=1}^N \frac{\omega_j}{2} (\hat{p}_j^2 + \hat{q}_j^2)
		+ E^{(0) b}
		+ \sum_{j=1}^N E^{(1) b}_j \hat{q}_j,
\end{align}
where $b = 1, \ldots, B$, and $B$ must be large enough that $\pi_{\vec{E}}$ is sufficiently flexible to describe $g$, but not so large that the optimization becomes too costly.
This is a natural form for this problem and allows us to easily use the same frequencies $\omega_j$ as in the full Hamiltonian $\hat{H}$ in Eq.~\eqref{eq:H-full}.
Additionally, because $\pi_{\vec{E}}$ may be generated from $\hat{h}_{\vec{E}}$ for any number of beads $P$, the same optimized parameters may be used for a range of $P$ values.

The primary constituent of the optimization loss function $\mathcal{L}(\vec{E})$ should be the relative entropy $D(\pi_{\vec{E}} \,||\, \abs{g}/Z_{||})$, but we can avoid some pitfalls in the optimization by adding extra terms.
One concern is with components $\pi_{\vec{E}}^b$ and $\pi_{\vec{E}}^{b'}$ that come too close to each other (that is, whose mean vectors $\vec{d}_{\vec{E}}^b$ and $\vec{d}_{\vec{E}}^{b'}$ become nearly the same).
When this happens, they are likely to follow the same trajectory during optimization, which is wasteful.
To encourage the components to steer clear of one another, we introduce a penalty term proportional to
\begin{align}
	\frac{2}{B (B-1)} \sum_{b=1}^{B-1} \sum_{b'=b+1}^B e^{-\frac{\abs{\vec{d}_{\vec{E}}^b - \vec{d}_{\vec{E}}^{b'}}^2}{2 \epsilon^2}},
\end{align}
where $\epsilon$ determines the stiffness of the repulsion.

Another potential issue is that some components $\pi_{\vec{E}}^b$ can fall out of favor and stop being considered important due to their low weights $w_{\vec{E}}^b$.
When this occurs, they cease being purposefully optimized and begin to aimlessly wander.
While this does allow them to explore highly unfavorable regions and therefore possibly cross barriers to find deeper minima, in our experience this is futile and should be suppressed.
To support those components which might find themselves at risk of becoming irrelevant, we penalize mixtures that have a low Shannon entropy\cite{shannon1948mathematical} of the weights using a term proportional to
\begin{align}
	1 + \frac{\sum_{b=1}^B w_{\vec{E}}^b \log w_{\vec{E}}^b}{\log{B}}.
\end{align}
This vanishes when all the weights are equal and grows to unity when a single component dominates the mixture.

The complete loss function that we employ is
\begin{align}
	\label{eq:loss}
	\mathcal{L}(\vec{E})
	&= D(\pi_{\vec{E}} \,||\, \abs{g}/Z_{||})
	\notag \\ &\quad
		+ C_1 \frac{2}{B (B-1)} \sum_{b=1}^{B-1} \sum_{b'=b+1}^B e^{-\frac{\abs{\vec{d}_{\vec{E}}^b - \vec{d}_{\vec{E}}^{b'}}^2}{2 \epsilon^2}}
	\notag \\ &\quad
		+ C_2 \left( 1 + \frac{\sum_{b=1}^B w_{\vec{E}}^b \log w_{\vec{E}}^b}{\log{B}} \right),
\end{align}
where $\epsilon$, $C_1$, and $C_2$ are tunable parameters, and only the first term has an associated statistical error.
To optimize the parameter tensor $\vec{E}$, we use SPSA, as it is both efficient and very simple to implement.\cite{spall1992multivariate,spall1998implementation}
The overall optimization algorithm is shown in Alg.~\ref{alg:optimization}.

\begin{figure}
\begin{algorithm}[H]
	\caption{GMD parameter optimization.}
	\label{alg:optimization}

	\begin{algorithmic}[0]
		\Require initial parameter tensor $\vec{E}$
		\Require $N_\mathrm{SPSA} \ge 1$
		\Require \Call{loss}{$\vec{E}$} estimates the loss function $\mathcal{L}$ in Eq.~\eqref{eq:loss}, returning a value and standard error
		\Require \Call{spsa}{\textsc{loss}, $k$, $\vec{E}$} is implemented as in Alg.~\ref{alg:spsa} in Sec.~\ref{sec:background-spsa}
		\Statex
		\State $\vec{E}_0 \gets \vec{E}$
		\Statex
		\For{$k \gets 1:N_\mathrm{SPSA}$}
			\State $\vec{E}_k \gets$ \Call{spsa}{\textsc{loss}, $k$, $\vec{E}_{k-1}$}
		\EndFor
		\Statex
		\State $\vec{E} \gets \vec{E}_{N_\mathrm{SPSA}}$
	\end{algorithmic}
\end{algorithm}
\end{figure}

Estimation of the relative entropy from Eq.~\eqref{eq:relative-entropy} is performed using the RqMC method with deterministic component selection outlined in Sec.~\ref{sec:methods-quasi}.\footnote{
	When the RqMC algorithm selects components, it tries to reduce the overall error in the estimate of the partition function.
	Ideally, it would be modified to attempt to reduce the error in the estimate of Eq.~\eqref{eq:relative-entropy} instead, but we find that this is not necessary.
}
When a single RqMC calculation is used to compute both terms of Eq.~\eqref{eq:relative-entropy}, evaluating the standard error of the mean is not as simple as for the partition function, because of the correlations between $\overline{\abs{f}}$ and $\overline{\log \abs{f}}$.
We may use the jackknife technique,\cite{young2015everything} but since there are several layers to RqMC, we must be careful about what exactly our statistical sample means are.
Recall that each of $\overline{\abs{f}}$ and $\overline{\log \abs{f}}$ is computed separately using the weights and component means, so a simple estimate of the sample mean is given by
\begin{align}
	& D\left( \overline{\abs{f^1}}, \ldots, \overline{\abs{f^B}}, \overline{\log \abs{f^1}}, \ldots, \overline{\log \abs{f^B}} \right)
	\notag \\
	&= \log\left[ \sum_{b=1}^B w^b \overline{\abs{f^b}} \right] - \sum_{b=1}^B w^b \overline{\log \abs{f^b}}.
\end{align}
Because each component mean $\overline{\abs{f^b}}$ and $\overline{\log{\abs{f^b}}}$ is computed as a statistical average over $N_\mathrm{S}$ independent random variables, we may apply jackknife to this function as usual.

\subsection{Parameter optimization with deformation}
\label{sec:methods-optimization-deformation}

To get around the circularity in Sec.~\ref{sec:methods-optimization} that arises from needing $\pi$ in order to determine $\pi$ in the general case, we propose an iterative scheme.
Recall that the function $g$ is derived from the vibronic Hamiltonian $\hat{H}$ in Eq.~\eqref{eq:H-full} as described in Sec.~\ref{sec:background-vibronic}.
We define the parameterized function $g_\nu$ (with $0 \le \nu \le 1$) to be similar to $g$, but instead derived from the parameterized Hamiltonian $\hat{H}_\nu$, where all the terms except the pure harmonic oscillator are scaled by $\nu$:
\begin{align}
	\hat{H}_\nu^{a a'}
	&= \nu \hat{H}^{a a'}
		+ (1 - \nu) \delta_{a a'} \hat{h}_\mathrm{o}.
\end{align}
When $\nu = 1$, we recover the original Hamiltonian $\hat{H} = \hat{H}_{\nu=1}$, so $g = g_{\nu=1}$; when $\nu = 0$, all off-diagonal terms vanish and the Hamiltonian $\hat{H}_{\nu=0}$ is trivial, so $g_{\nu=0}$ is proportional to a GMD.
This allows us to smoothly interpolate between easily solvable models and chemically interesting ones.
Thus, we can use $g_\nu$ to guide the optimization of $\pi$ towards $g$.
For each $g_\nu$, we introduce the normalized and sign-free distribution
\begin{align}
	\frac{\abs{g_\nu(\vec{R})}}{Z_{\nu ||}}
	&= \frac{\abs{g_\nu(\vec{R})}}{\int\! \dd{\vec{R}} \abs{g_\nu(\vec{R})}}.
\end{align}

In practice, we must choose a monotonically increasing sequence $\nu_n$, where $\nu_1 > 0$ and $\nu_{N_\nu} = 1$, such that each step $\Delta \nu_n = \nu_n - \nu_{n-1}$ is small enough to obtain accurate estimates, yet large enough to make rapid progress toward the desired distribution.
To select the $\nu_n$, we use the function \textsc{step\_nu} in Alg.~\ref{alg:step-nu}, which is described in more detail in Appendix~\ref{sec:step-nu}.
At the final iteration $n = N_\nu$, the target distribution is $\abs{g} / Z_{||}$ itself, so the optimized GMD should be adequate for evaluating $Z = \ev{f}_{\pi}$.

To initialize the coefficients $E^{(0) b}$ and $E^{(1) b}_j$ making up the tensor $\vec{E}$ in Eq.~\eqref{eq:gmd-hamiltonian}, we set them all to zero; this leaves us with only the pure harmonic oscillator terms:
\begin{align}
	\hat{h}_{\vec{E}=\vec{0}}^b
	&= \sum_{j=1}^N \frac{\omega_j}{2} (\hat{p}_j^2 + \hat{q}_j^2)
	= \hat{h}_\mathrm{o}.
\end{align}
Hence, the initial distribution $\pi_{\vec{E}=\vec{0}}$ resembles $g_{\nu=0}$, but potentially with a different number of components.

For each $n = 1, \ldots, N_\nu$, we perturb $\vec{E}$ using the optimization routine from Alg.~\ref{alg:optimization} to minimize the loss function
\begin{align}
	\label{eq:loss-deform}
	\mathcal{L}_\nu(\vec{E})
	&= D(\pi_{\vec{E}} \,||\, \abs{g_\nu}/Z_{\nu ||})
	\notag \\ &\quad
		+ C_1 \frac{2}{B (B-1)} \sum_{b=1}^{B-1} \sum_{b'=b+1}^B e^{-\frac{\abs{\vec{d}_{\vec{E}}^b - \vec{d}_{\vec{E}}^{b'}}^2}{2 \epsilon^2}}
	\notag \\ &\quad
		+ C_2 \left( 1 + \frac{\sum_{b=1}^B w_{\vec{E}}^b \log w_{\vec{E}}^b}{\log{B}} \right),
\end{align}
where the second and third terms are the same as in Eq.~\eqref{eq:loss}.
We take the best of these perturbed distributions $\pi_{\vec{E}}$ and promote it to be the starting distribution for the next iteration.
The overall algorithm for optimization with deformation can be seen in Alg.~\ref{alg:optimization-deformation}.
Note that we use $N_\mathrm{W}$ independent ``walkers'' and take the result from the best performing one (in the sense of minimizing the loss function); because they are independent, they may be executed in parallel.

\begin{figure}
\begin{algorithm}[H]
	\caption{Step size selection for $\nu$.}
	\label{alg:step-nu}

	\begin{algorithmic}[0]
		\Require \Call{loss$_\nu$}{$\vec{E}$} estimates the loss function $\mathcal{L}_\nu$ in Eq.~\eqref{eq:loss-deform}, returning a value and standard error
		\Statex
		\Function{step\_nu}{$\nu_\mathrm{prev}$, $\Delta \nu_\mathrm{prev}$, $\vec{E}$}
			\State $\Delta \nu_\mathrm{min} \gets 10^{-3}$
			\If{$\nu_\mathrm{prev} > 1 - 10^{-1} - 10^{-2}$}
				\State $\Delta \nu_\mathrm{max} \gets 10^{-2}$
			\Else
				\State $\Delta \nu_\mathrm{max} \gets 10^{-1}$
			\EndIf
			\Statex
			\State $\Delta \nu \gets$ \Call{min}{$\Delta \nu_\mathrm{max}$, $2 \Delta \nu_\mathrm{prev}$}
			\Statex
			\For{$i \gets 1:8$}
				\State $\nu_\mathrm{new} \gets$ \Call{min}{1, $\nu_\mathrm{prev} + \Delta \nu$}
				\State $\ell, \sigma \gets$ \Call{loss$_{\nu_\mathrm{new}}$}{$\vec{E}$}
				\Statex
				\If{$\ell < 1$}
					\State $\Delta \nu_\mathrm{min} \gets \Delta \nu$
				\ElsIf{$\ell > 2$}
					\State $\Delta \nu_\mathrm{max} \gets \Delta \nu$
				\Else
					\State \Call{break}{}
				\EndIf
				\Statex
				\State $\Delta \nu \gets (\Delta \nu_\mathrm{min} + \Delta \nu_\mathrm{max})/2$
			\EndFor
			\Statex
			\State \Call{return}{$\Delta \nu$}
		\EndFunction
	\end{algorithmic}
\end{algorithm}
\end{figure}

\begin{figure}
\begin{algorithm}[H]
	\caption{GMD parameter optimization with deformation.}
	\label{alg:optimization-deformation}

	\begin{algorithmic}[0]
		\Require $N_\mathrm{SPSA} \ge 1$
		\Require $N_\mathrm{W} \ge 1$
		\Require \Call{loss$_\nu$}{$\vec{E}$} estimates the loss function $\mathcal{L}_\nu$ in Eq.~\eqref{eq:loss-deform}, returning a value and standard error
		\Require \Call{spsa}{\textsc{loss}, $k$, $\vec{E}$} is implemented as in Alg.~\ref{alg:spsa} in Sec.~\ref{sec:background-spsa}
		\Require \Call{step\_nu}{$\nu_\mathrm{prev}$, $\Delta \nu_\mathrm{prev}$, $\vec{E}$} is implemented as in Alg.~\ref{alg:step-nu}
		\Statex
		\For{$b \gets 1:B$}
			\Comment{Initialization.}
			\State $E^{(0) b} \gets 0$

			\For{$j \gets 1:N$}
				\State $E^{(1) b}_j \gets 0$
			\EndFor
		\EndFor
		\Statex
		\State $\nu \gets 0$
		\State $\Delta \nu \gets 10^{-1}$
		\Statex
		\While{$\nu < 1$}
			\Comment{Iteration.}
			\State $\Delta \nu \gets$ \Call{step\_nu}{$\nu$, $\Delta \nu$, $\vec{E}$}
			\Comment{$\nu$ selection.}
			\State $\nu \gets$ \Call{min}{1, $\nu + \Delta \nu$}
			\Statex
			\For{$i \gets 1:N_\mathrm{W}$}
				\Comment{Parameter update.}
				\State $\vec{E}_{i 0} \gets \vec{E}$

				\For{$k \gets 1:N_\mathrm{SPSA}$}
					\State $\vec{E}_{i k} \gets$ \Call{spsa}{\textsc{loss}$_\nu$, $k$, $\vec{E}_{i, k-1}$}
					\State $\ell_{i k}, \sigma_{i k} \gets$ \Call{loss$_\nu$}{$\vec{E}_{i k}$}
				\EndFor
			\EndFor
			\Statex
			\State $i, k \gets \Call{argmin}{\vec{\ell} + \vec{\sigma}}$
			\State $\vec{E} \gets \vec{E}_{i k}$
		\EndWhile
	\end{algorithmic}
\end{algorithm}
\end{figure}

\clearpage

\section{Results}
\label{sec:results}

In this section, we apply the methods from Sec.~\ref{sec:methods} to model systems.
All the systems that we consider have two diabatic surfaces and two spatial degrees of freedom, with the Hamiltonian
\begin{align}
	\label{eq:model-system}
	\hat{H}
	&= \begin{pmatrix}
				E^1 + \hat{h}_\mathrm{o} + \lambda \hat{q}_1 & 0 \\
				0 & E^2 + \hat{h}_\mathrm{o} - \lambda \hat{q}_1 \\
			\end{pmatrix}
		+ \gamma \begin{pmatrix}
				0 & \hat{q}_2 \\
				\hat{q}_2 & 0 \\
			\end{pmatrix},
\end{align}
which has six parameters.
We use an inverse temperature of $\beta = 38.7$, which is approximately \SI{300}{\kelvin} if the parameter values are interpreted in \si{\electronvolt}.
Where applicable, we use $N_\mathrm{S} = 100$ to ensure sufficiently many randomized sequences for accurate error bars, and $N_\mathrm{boot} = 64$ to obtain reasonable initial estimates.
The sole exception to this is during GMD parameter optimization, when quality estimates are not required, in which case we use $N_\mathrm{S} = 50$ and $N_\mathrm{boot} = 4$.
The algorithms are implemented using VibronicToolkit.\cite{vibronictoolkitjl}

\subsection{Deterministic component selection}
\label{sec:results-component}

We start with a contrived example that highlights the ability of the deterministic component selection algorithm described in Sec.~\ref{sec:methods-component} to reduce the impact of outliers.
We use the model Hamiltonian Eq.~\eqref{eq:model-system} with the parameters given in Tab.~\ref{tab:displaced-mod-parameters}, which are similar to those of the Displaced model of Ref.~\onlinecite{raymond2018path} with the $\gamma_2$ parameter.

\begin{table}
	\caption{
		Parameters of the model Hamiltonian in Eq.~\eqref{eq:model-system} for the modified Displaced $\gamma_2$ system.
	}
	\label{tab:displaced-mod-parameters}
	\hfill{}
	\begin{tabular}{c S[table-format=1.3]}
		\toprule
		Parameter & {Value} \\
		\midrule
		$E^1$ & 0.1 \\
		$E^2$ & 0.275 \\
		$\omega_1$ & 0.02 \\
		$\omega_2$ & 0.04 \\
		\botrule
	\end{tabular}
	\hfill{}
	\begin{tabular}{c S[table-format=1.3]}
		\toprule
		Parameter & {Value} \\
		\midrule
		$\lambda$ & 0.075 \\
		$\gamma$ & 0.05 \\
		\botrule
	\end{tabular}
	\hfill{}
\end{table}

The coupling between the surfaces is not very strong, so we expect the basic sampling approach to work well, and we use the diagonal portion of $\hat{H}$ to construct the two-component sampling distribution $\pi_0$.
At a temperature of $\beta = 38.7$, the component weights are $w^{b=1} \approx \num{0.99886}$ and $w^{b=2} \approx \num{0.00114}$.
Thus, out of every \num{1000} samples, approximately one should be drawn from the second component.

\begin{figure}
	\includegraphics{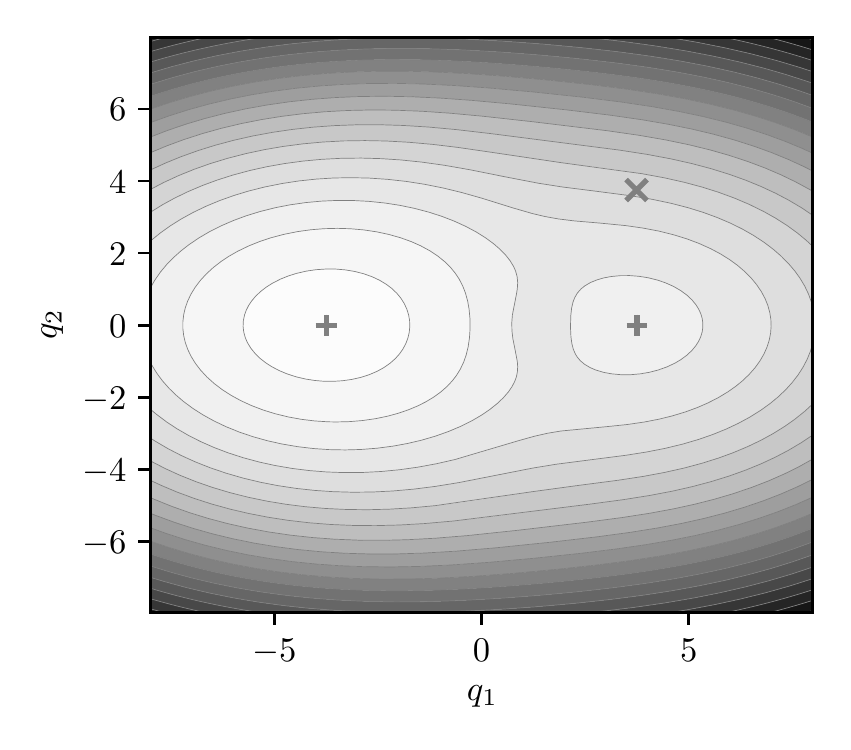}
	\caption{
		Ground state potential energy surface of the modified Displaced $\gamma_2$ model.
		The locations of the GMD component minima are marked with $+$ signs.
		The position of the forced outlier from Eq.~\eqref{eq:outlier-position} is marked with a $\times$ sign.
	}
	\label{fig:displaced-mod-pes-outlier}
\end{figure}

What makes this example particularly artificial is that we force the second point drawn from the second component to be an extreme outlier.
Recall from Ref.~\onlinecite{raymond2018path} that each path is sampled from the component $\pi_0^b$ using uncoupled coordinates $y_{j \lambda}$ (where $j = 1, \ldots, N$ and $\lambda = 1, \ldots, P$), each with a standard deviation $\sigma^b_{j \lambda}$.
For the outlier point, we choose the coordinates
\begin{align}
	\label{eq:outlier-position}
	y_{j \lambda}
	&= \begin{dcases*}
			6 \sigma^{b=2}_{j \lambda} & if $j = 2$ and $\lambda = 1$ \\
			0.01 \sigma^{b=2}_{j \lambda} & otherwise,
		\end{dcases*}
\end{align}
which place the centroid mode of the path quite far from its mean in the second spatial coordinate, as shown in Fig.~\ref{fig:displaced-mod-pes-outlier}.
Since the cdf of a univariate Gaussian is (with $\erf$ being the error function)
\begin{align}
	\Phi_\sigma(x)
	&= \frac{1}{2} \left( 1 + \erf{\frac{x}{\sqrt{2 \sigma^2}}} \right),
\end{align}
the total probability of either $x < -n \sigma$ or $x > n \sigma$ is
\begin{align}
	1 - \erf{\frac{n}{\sqrt{2}}}.
\end{align}
Hence, when $P = 16$, the probability of sampling a point that's at least as unlikely in all directions is
\begin{align}
	\left( 1 - \erf{\frac{6}{\sqrt{2}}} \right)
	\left( 1 - \erf{\frac{0.01}{\sqrt{2}}} \right)^{31}
	\approx \num{1.539276e-9}.
\end{align}
Admittedly, this is not a very probable event.
However, when taking huge numbers of samples over many calculations, very unlikely things are bound to happen occasionally.

\begin{figure*}
	\includegraphics{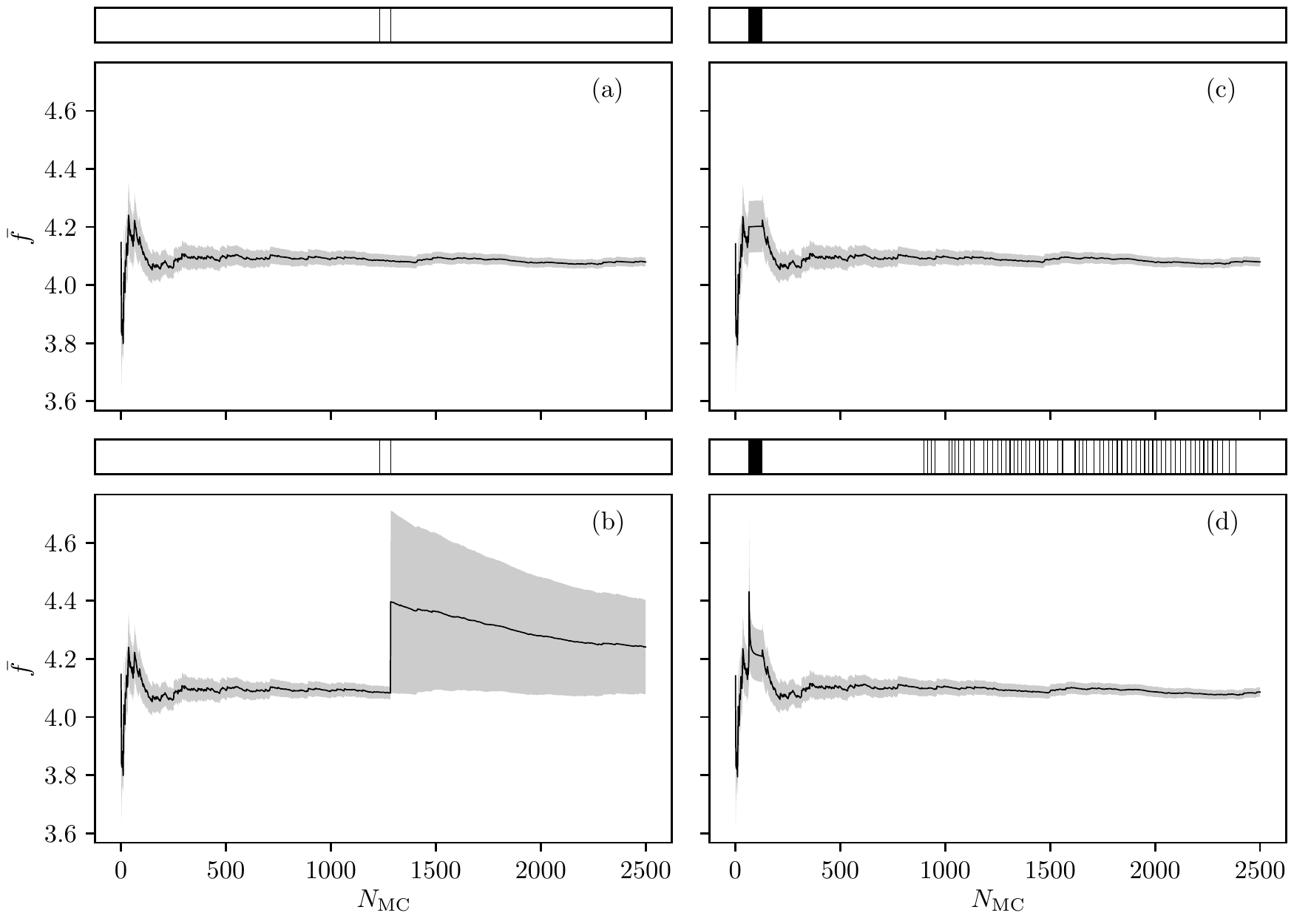}
	\caption{
		Convergence of $\bar{f}$ for the modified Displaced $\gamma_2$ model with $P = 16$ and (a) stochastic component selection with no outlier; (b) stochastic component selection with an outlier; (c) deterministic component selection with no outlier; (d) deterministic component selection with an outlier.
		The shaded area reflects the instantaneous error estimates.
		The vertical marks above the plots indicate steps at which the sample was drawn from the second component; in (c) and (d), this occurs $N_\mathrm{boot}$ times during bootstrapping.
	}
	\label{fig:displaced-mod-convergence}
\end{figure*}

We compare the convergence of $Z \approx \bar{f}$ between four scenarios.
In Fig.~\ref{fig:displaced-mod-convergence}(a), we have the simplest case: familiar stochastic component selection, and no forced outlier; here, the mean stabilizes relatively quickly, and the second component is sampled twice, as expected.
The forced outlier is clearly visible in Fig.~\ref{fig:displaced-mod-convergence}(b), where the stochastic algorithm accepts it as yet another point, greatly changing the mean and increasing the standard error.
Fig.~\ref{fig:displaced-mod-convergence}(c) features the deterministic algorithm and looks very similar to (a), except the second component is sampled multiple times during the bootstrap phase.
Finally, in Fig.~\ref{fig:displaced-mod-convergence}(d), the outlier is found to have a strong impact during bootstrapping, but this is quickly quelled; additionally, in the remainder of the calculation, the outlier causes more samples to be drawn from the second component.
The final result is similar in all scenarios, except the stochastic algorithm with a forced outlier in Fig.~\ref{fig:displaced-mod-convergence}(b), which fares poorly.

\begin{figure}
	\includegraphics{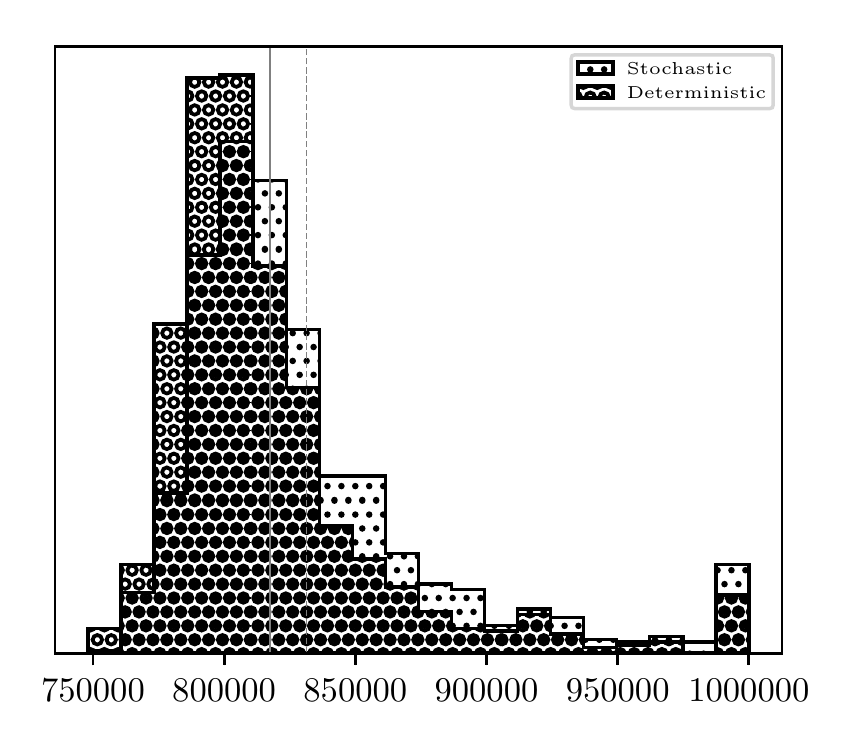}
	\caption{
		Distributions of the numbers of samples required to obtain a standard error of the mean of \num{e-3} for the modified Displaced $\gamma_2$ model.
		To avoid very wide histograms, the right-most bins collect all the points with more than \num{e6} samples.
		The vertical lines indicate the means, dashed for stochastic and solid for deterministic.
	}
	\label{fig:displaced-mod-histogram}
\end{figure}

We also consider a more legitimate example, which doesn't require any elaborate setup of the sampling.
The same model system and sampling distribution are used, but without any forced outliers.
We run \num{1000} calculations with the stochastic component selection algorithm, and \num{1000} more with the deterministic one.
These calculations are stopped when the standard error of the mean reaches \num{e-3}, and the number of samples (including the $N_\mathrm{boot} = 64$ bootstrap samples per component for the deterministic version) is recorded.
The numbers of samples are then binned to produce the histograms in Fig.~\ref{fig:displaced-mod-histogram}, where it is clear that the deterministic method tends to require slightly fewer samples to achieve the same level of error.
The reduction in the mean number of samples is about \SI{1.7}{\percent}.

\subsection{Randomized quasi-Monte Carlo}
\label{sec:results-quasi}

\begin{figure}
	\includegraphics{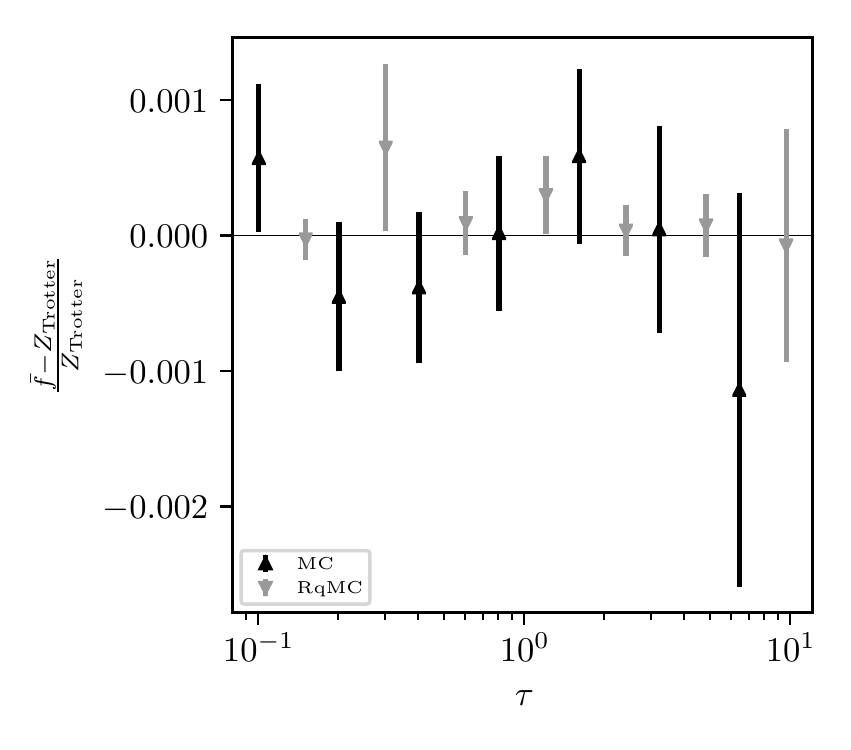}
	\caption{
		Comparison of RqMC with MC for the convergence of $\bar{f}$ with imaginary time step size $\tau$.
		The data are plotted relative to the exact result $Z_\mathrm{Trotter}$.
		Values of $\tau$ are staggered between the two methods for better visibility ($P = 6, 12, 24, 48, 96, 192, 384$ for MC and $P = 4, 8, 16, 32, 64, 128, 256$ for RqMC).
	}
	\label{fig:displaced-mod-Z}
\end{figure}

In order to evaluate the efficiency of RqMC compared to MC, we use the same modified Displaced $\gamma_2$ model system in Tab.~\ref{tab:displaced-mod-parameters} as in Sec.~\ref{sec:results-component}, along with the same sampling distribution $\pi_0$.
The approach for the comparison is simple: for several values of $\tau$, we estimate $Z \approx \bar{f}$ from a fixed number of samples ($N_\mathrm{MC} = \num{e5}$), and compare it to the exact result $Z_\mathrm{Trotter}$, which includes the systematic error due to the Trotter factorization.
It is clear from Fig.~\ref{fig:displaced-mod-Z} that RqMC tends to result in smaller error bars than MC for this system, even though we use the deterministic component selection algorithm for MC.

\begin{figure}
	\includegraphics{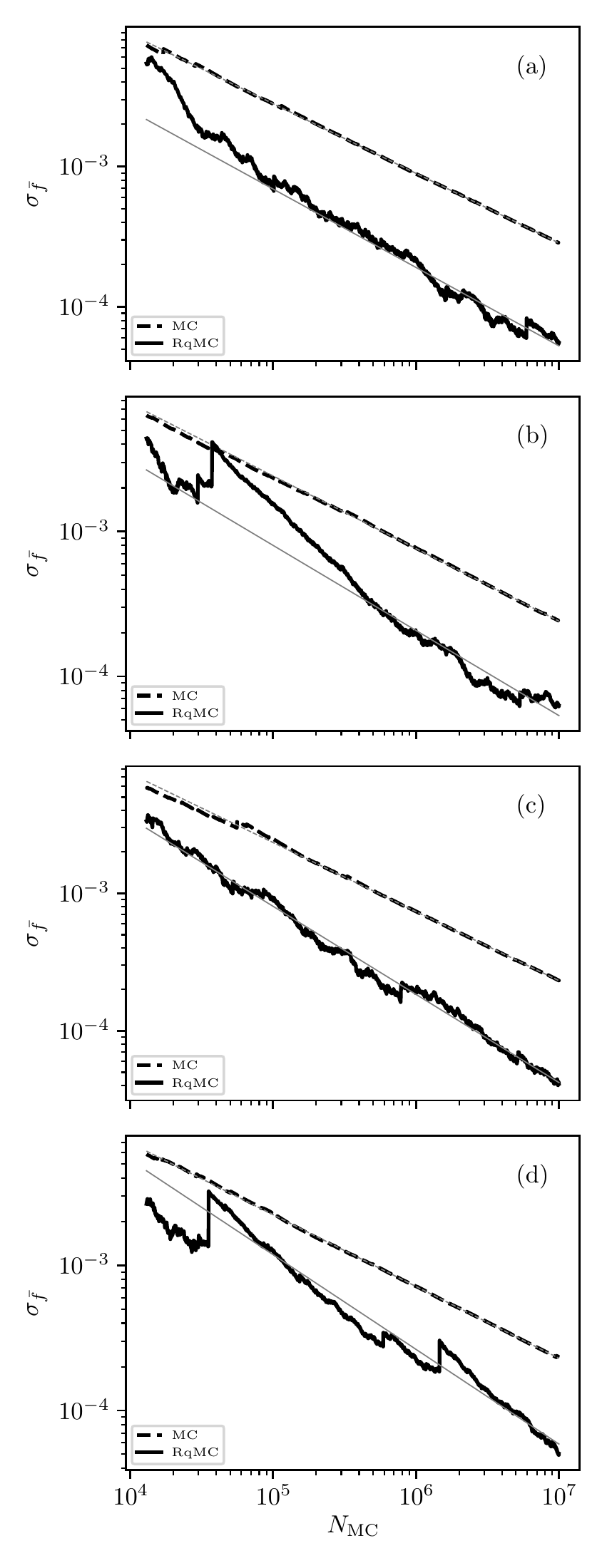}
	\caption{
		Scaling of the estimated error in $\bar{f}$ with the number of samples for $P = $ (a) 16, (b) 32, (c) 64, (d) 128.
		In grey are linear least squares fits whose slopes are given in Tab.~\ref{tab:displaced-mod-error-slope}.
	}
	\label{fig:displaced-mod-error-scaling}
\end{figure}

We also present in Fig.~\ref{fig:displaced-mod-error-scaling} a comparison of the error scaling with the number of samples.
We compare the scaling for several $P$ values (see Tab.~\ref{tab:displaced-mod-error-slope}), finding it extremely consistent for MC, where the slope is always near $-1/2$, indicating an asymptotic error scaling proportional to $N_\mathrm{MC}^{-\frac{1}{2}}$, which is precisely what one expects for Monte Carlo.
In the RqMC case, the raw data are not as smooth (on account of being computed from just $N_S$ points instead of $N_\mathrm{MC}$), so the variability in the slopes is greater; still, we find that they are more negative than their MC counterparts, implying faster error reduction.

\begin{table}
	\caption{
		Slopes of linear fits in Fig.~\ref{fig:displaced-mod-error-scaling}.
	}
	\label{tab:displaced-mod-error-slope}
	\begin{tabular}{S[table-format=3] S[table-format=+1.4] S[table-format=+1.5]}
		\toprule
		{$P$} & {Slope (MC)} & {Slope (RqMC)} \\
		\midrule
		16 & -0.49647 & -0.55876 \\
		32 & -0.50114 & -0.58758 \\
		64 & -0.50190 & -0.64040 \\
		128 & -0.49453 & -0.65234 \\
		\botrule
	\end{tabular}
\end{table}

\subsection{Parameter optimization}
\label{sec:results-optimization}

The GMD parameter optimization algorithm described in Sec.~\ref{sec:methods-optimization} is tested here using a simple model system derived from the Displaced model of Ref.~\onlinecite{raymond2018path} with the $\gamma_6$ parameter by reducing the magnitude of the non-frequency parameters.
The resulting system parameters are given in Tab.~\ref{tab:displaced-simple-parameters}, while those for the algorithm are in Tab.~\ref{tab:optimization-parameters}.

\begin{table}[p]
	\caption{
		Parameters of the model Hamiltonian in Eq.~\eqref{eq:model-system} for the weakened Displaced $\gamma_6$ system.
	}
	\label{tab:displaced-simple-parameters}
	\hfill{}
	\begin{tabular}{c S[table-format=1.4]}
		\toprule
		Parameter & {Value} \\
		\midrule
		$E^1$ & 0.02 \\
		$E^2$ & 0.04 \\
		$\omega_1$ & 0.02 \\
		$\omega_2$ & 0.04 \\
		\botrule
	\end{tabular}
	\hfill{}
	\begin{tabular}{c S[table-format=1.4]}
		\toprule
		Parameter & {Value} \\
		\midrule
		$\lambda$ & 0.01 \\
		$\gamma$ & 0.05 \\
		\botrule
	\end{tabular}
	\hfill{}
\end{table}

\begin{figure}
	\includegraphics{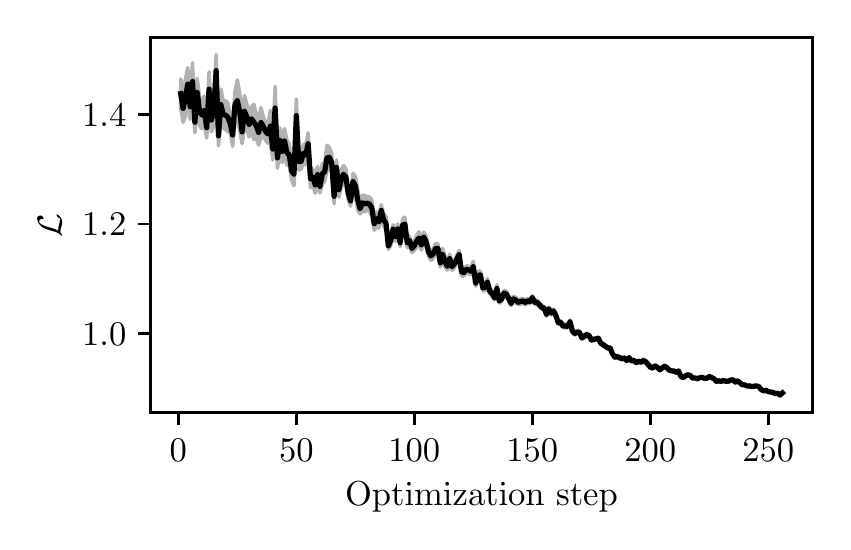}
	\caption{
		The loss function $\mathcal{L}$ at each step of the GMD optimization for the weakened Displaced $\gamma_6$ model.
		The uncertainty of the loss function estimates is indicated by the shaded area.
	}
	\label{fig:loss-displaced-6-weak}
\end{figure}

\begin{figure}
	\includegraphics{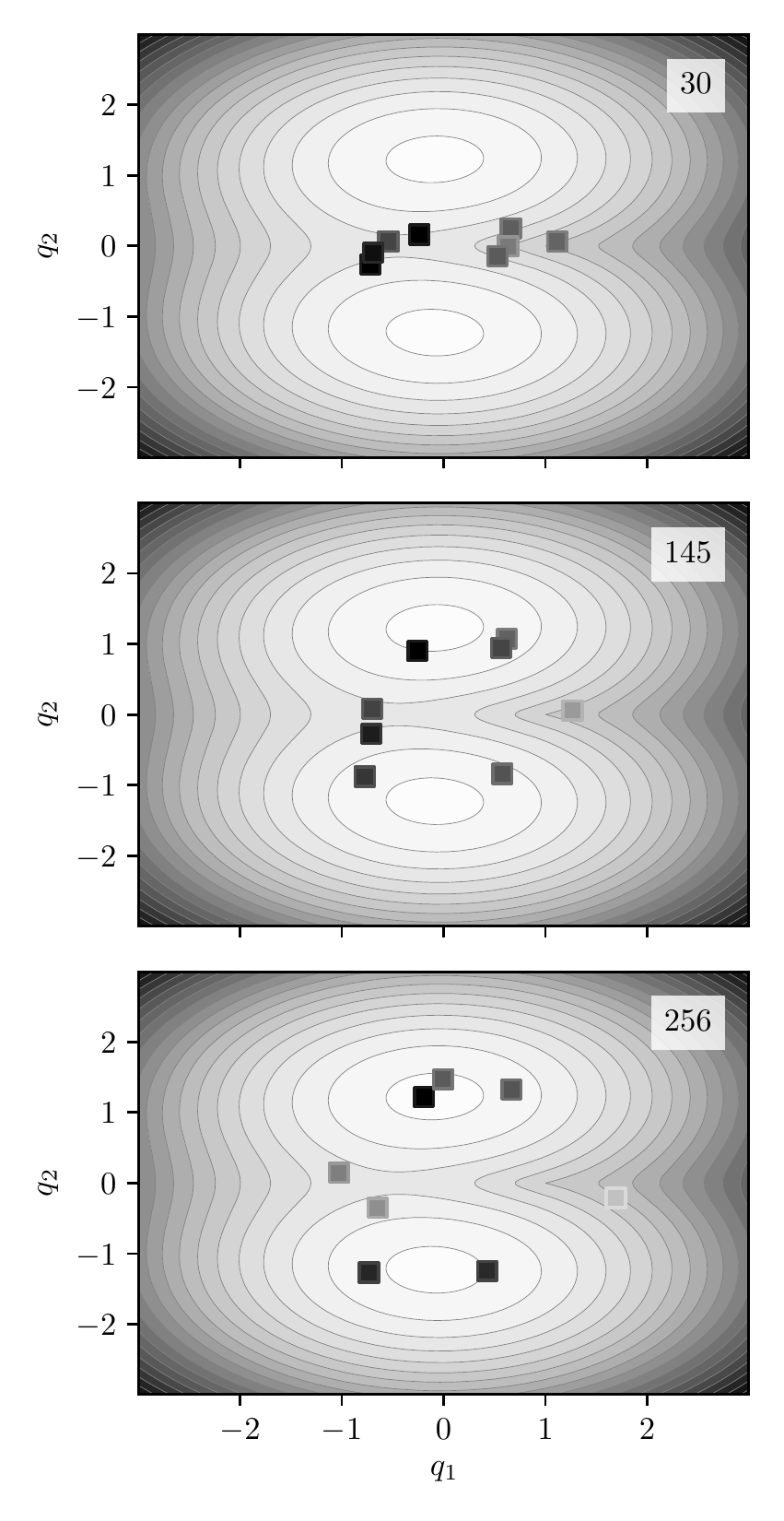}
	\caption{
		Snapshots of the GMD optimization for the weakened Displaced $\gamma_6$ model.
		The number in each panel indicates the optimization step.
		The background is the ground state potential energy surface of $\hat{H}$.
		The minimum of each GMD component is represented by a square, colored according to the relative component weights.
	}
	\label{fig:pes-displaced-6-weak}
\end{figure}

The sampling distribution is initially determined from the diagonal portion of the Hamiltonian, but with each GMD component replicated to make up a total of $B = 8$ components.
It is then optimized using Alg.~\ref{alg:optimization} to better match the true distribution arising from the entire Hamiltonian.
The progress of this optimization can be seen in Fig.~\ref{fig:loss-displaced-6-weak} and Fig.~\ref{fig:pes-displaced-6-weak}; the former shows the loss function, while the latter displays the motion of the GMD components.
As the oscillators find their way to the minima of the potential energy surface, the loss function steadily decreases.
The standard error of the loss function estimates also lessens as the optimized distribution gets closer to the desired distribution.
The components' journey is a fairly boring one: about half-way through, they get near the mimima, and towards the end, the ones closest to the minima have the largest weights.

\begin{table}
	\caption{
		Parameters for GMD optimization.
	}
	\label{tab:optimization-parameters}
	\begin{tabular}{c S[table-format=5.4] l}
		\toprule
		Parameter & {Value} & Description \\
		\midrule
		$P$ & 64 & \# of beads \\
		$B$ & 8 & \# of components \\
		$a$ & 0.002 & SPSA gain factor \\
		$N_\mathrm{SPSA}$ & 256 & \# of SPSA steps \\
		$N_\mathrm{MC}$ & 2000 & \# of samples per $\mathcal{L}$ estimation \\
		$\epsilon$ & 3 & component repulsion stiffness \\
		$C_1$ & 1 & prefactor for repulsion term \\
		$C_2$ & 1.5 & prefactor for weight balancing term \\
		\botrule
	\end{tabular}
\end{table}

\begin{figure}
	\includegraphics{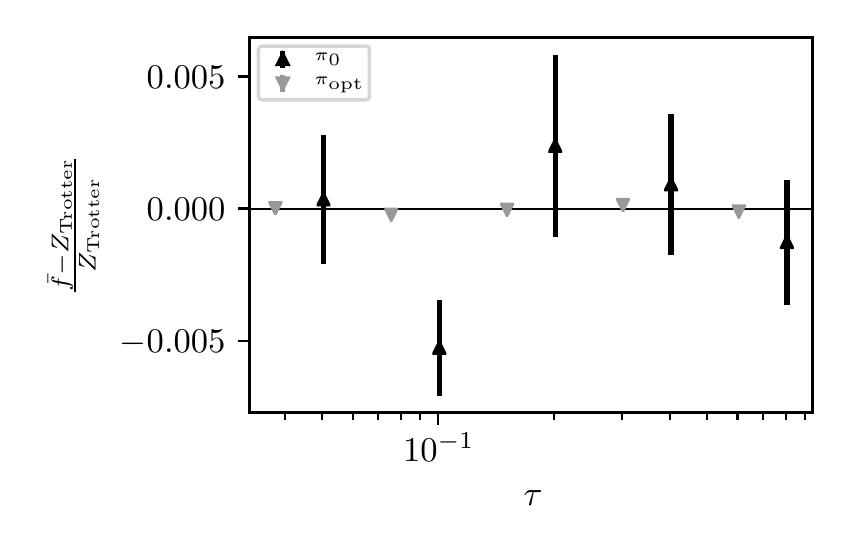}
	\caption{
		Comparison of sampling distributions for the convergence of $\bar{f}$ with imaginary time step size $\tau$ for the weakened Displaced $\gamma_6$ model.
		The data are plotted relative to the exact result $Z_\mathrm{Trotter}$.
		Values of $\tau$ are staggered between the two distributions for better visibility ($P = 48, 96, 192, 384, 768$ for $\pi_0$ and $P = 64, 128, 256, 512, 1024$ for $\pi_\mathrm{opt}$).
	}
	\label{fig:optimized-convergence}
\end{figure}

To demonstrate that the optimized distribution $\pi_\mathrm{opt}$ is an improvement over the simple diagonal sampling distribution $\pi_0$, we compare estimates of the partition function $Z \approx \bar{f}$ at various values of $P$ from $N_\mathrm{MC} = \num{e5}$ samples.
As evident in Fig.~\ref{fig:optimized-convergence}, $\pi_0$ already results in relatively small deviations from the exact values, but $\pi_\mathrm{opt}$ is undeniably the better choice.

\vspace{5em}

\subsection{Parameter optimization with deformation}
\label{sec:results-optimization-deformation}

To evaluate the algorithm for parameter optimization with deformation, we use one system with weak coupling and two with strong coupling.
The sampling distribution for each system is optimized using the procedure described in Sec.~\ref{sec:methods-optimization-deformation} and the parameters in Tab.~\ref{tab:optimization-deformed-parameters}.

\begin{table}
	\caption{
		Parameters for GMD optimization with deformation.
	}
	\label{tab:optimization-deformed-parameters}
	\begin{tabular}{c S[table-format=5.4] l}
		\toprule
		Parameter & {Value} & Description \\
		\midrule
		$P$ & 64 & \# of beads \\
		$B$ & 8 & \# of components \\
		$a$ & 0.0005 & SPSA gain factor \\
		$N_\mathrm{SPSA}$ & 64 & \# of SPSA steps \\
		$N_\mathrm{W}$ & 10 & \# of SPSA walkers \\
		$N_\mathrm{MC}$ & 2000 & \# of samples per $\mathcal{L}$ estimation \\
		$\epsilon$ & 3 & component repulsion stiffness \\
		$C_1$ & 1 & prefactor for repulsion term \\
		$C_2$ & 1.5 & prefactor for weight balancing term \\
		\botrule
	\end{tabular}
\end{table}

To evaluate the quality of the optimized distribution $\pi_\mathrm{opt}$ for each system, we use it with the RqMC method to estimate the partition function $Z \approx \bar{f}$ at various values of $P$.
For comparison, we also provide results computed using the best hand-picked sampling distribution of Ref.~\onlinecite{raymond2018path} for the system in question, which we call $\pi_\mathrm{best}$.
For both distributions, we use $N_\mathrm{MC} = \num{e5}$.

\subsubsection{Displaced $\gamma_2$}

\begin{figure}
	\includegraphics{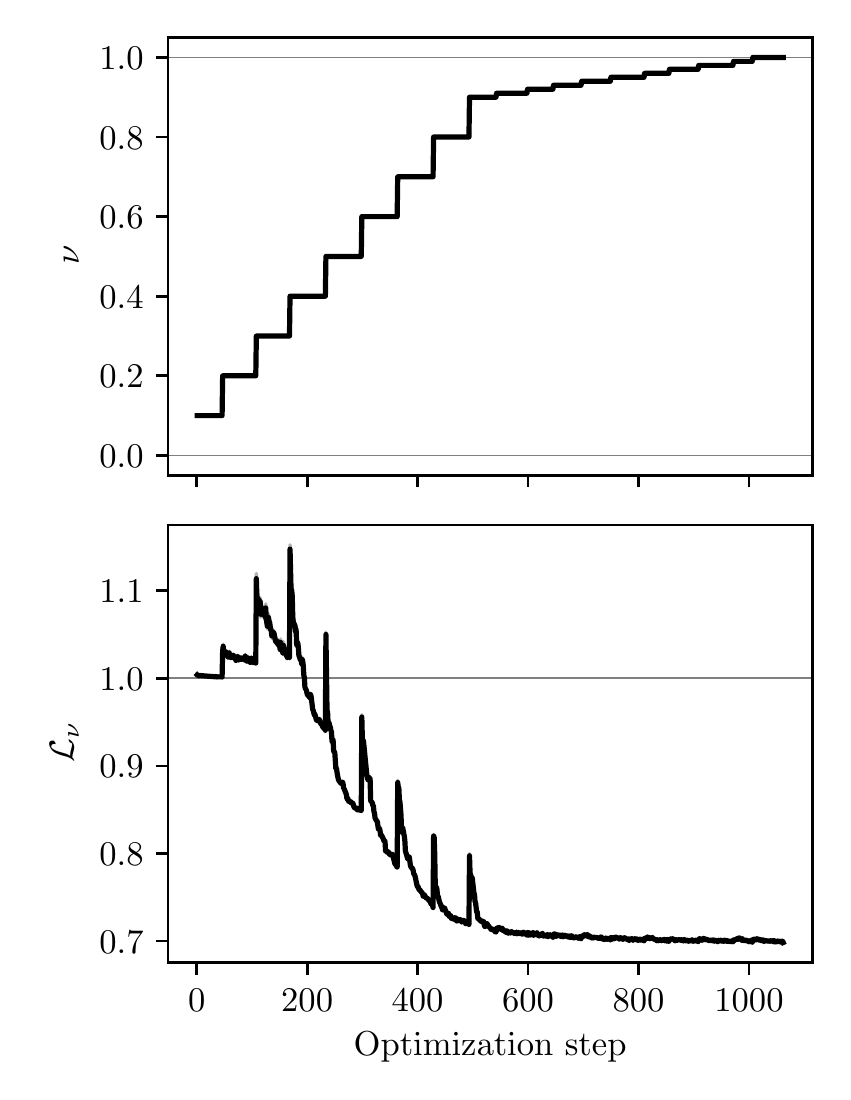}
	\caption{
		The value of $\nu$ (top panel) and the loss function $\mathcal{L}_\nu$ (bottom panel) at each step of the GMD optimization for the Displaced $\gamma_2$ model.
		In the top panel, the horizontal lines indicate the bounds on $\nu$.
		In the bottom panel, the horizontal line marks the lower bound on the desired loss function values, and the uncertainty of the loss function estimates is indicated by the shaded area.
	}
	\label{fig:loss-displaced-2}
\end{figure}

\begin{figure}
	\includegraphics{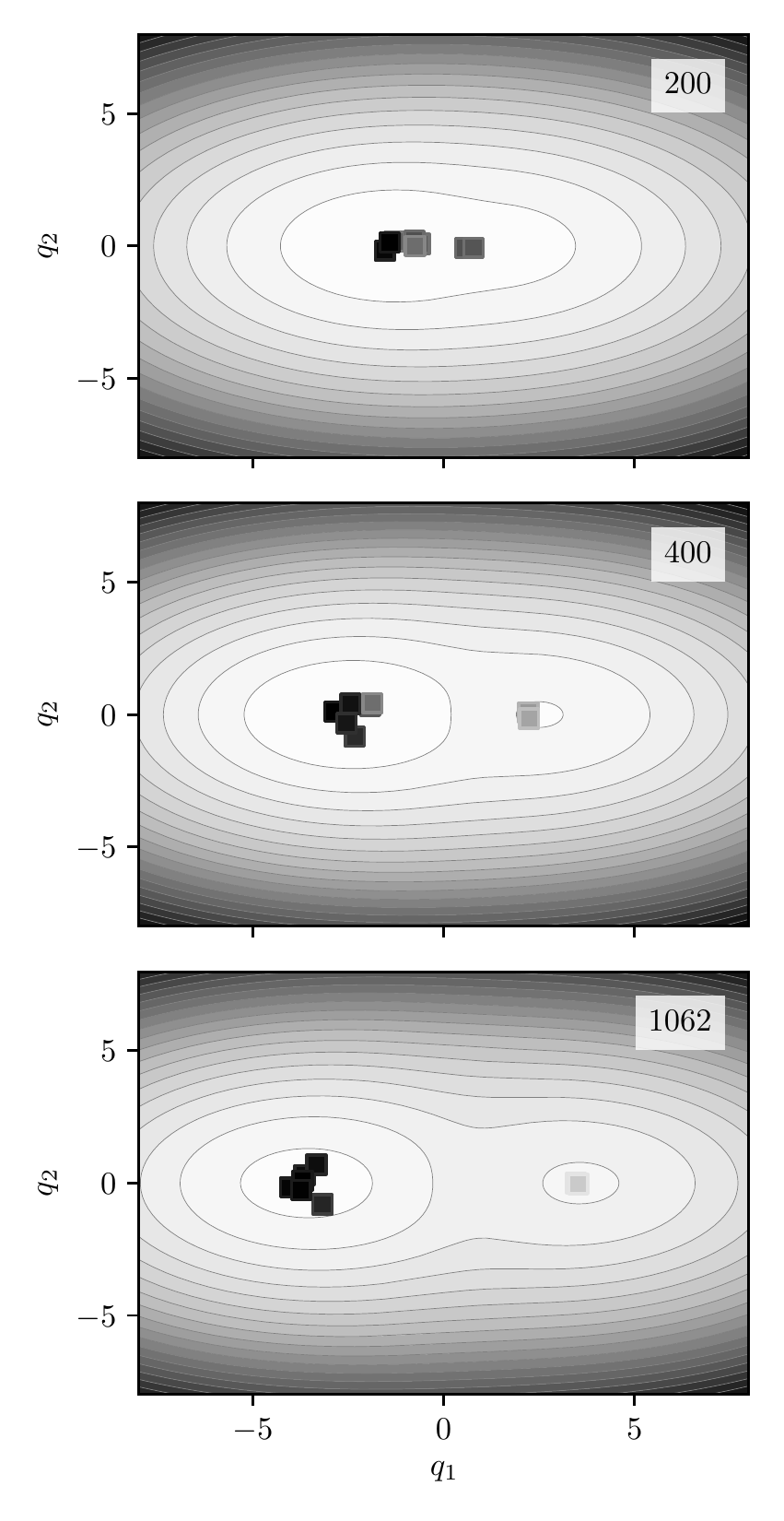}
	\caption{
		Snapshots of the GMD optimization for the Displaced $\gamma_2$ model.
		The number in each panel indicates the optimization step.
		The backgrounds are the ground state potential energy surfaces of $\hat{H}_\nu$.
		The minimum of each GMD component is represented by a square, colored according to the relative component weights.
	}
	\label{fig:pes-displaced-2}
\end{figure}

We start with a weakly-coupled model system: the $\gamma_2$ version of the Displaced model from Ref.~\onlinecite{raymond2018path}, which has the parameters given in Tab.~\ref{tab:displaced-2-parameters}.
The optimization progress is shown in Fig.~\ref{fig:loss-displaced-2} and Fig.~\ref{fig:pes-displaced-2}.
Because this is a fairly simple system, the loss function never takes on large values and the optimization algorithm always makes the largest $\nu$ step possible, which makes for a quick optimization.
As the Hamiltonian is deformed, the oscillators track the minima of the ground state potential energy surface.
The end result has most of the components located in the deeper minimum and having larger weights, and two components in the shallower minimum with smaller weights.

\begin{figure}
	\includegraphics{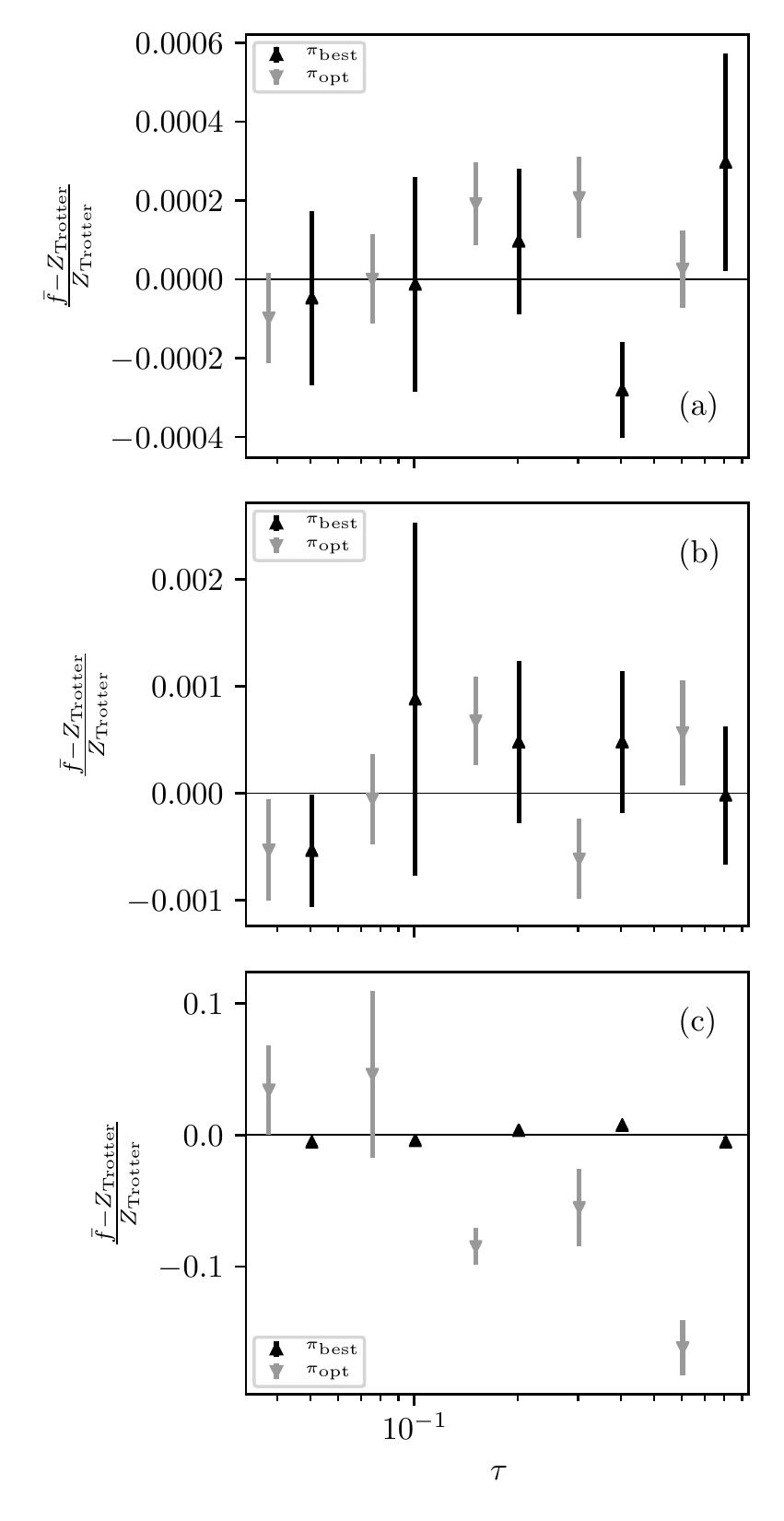}
	\caption{
		Comparison of sampling distributions for the convergence of $\bar{f}$ with imaginary time step size $\tau$ for (a) Displaced $\gamma_2$, (b) Displaced $\gamma_6$, and (c) Jahn--Teller $\lambda_6$.
		The data are plotted relative to the exact result $Z_\mathrm{Trotter}$.
		Values of $\tau$ are staggered between the two distributions for better visibility ($P = 48, 96, 192, 384, 768$ for $\pi_\mathrm{best}$ and $P = 64, 128, 256, 512, 1024$ for $\pi_\mathrm{opt}$).
	}
	\label{fig:optimized-deformed-convergence}
\end{figure}

As the reference distribution $\pi_\mathrm{best}$ for the sampling efficiency comparison, we simply use the one built from the diagonal portion of the Hamiltonian, which Ref.~\onlinecite{raymond2018path} calls $\rho_0$.
The results shown in Fig.~\ref{fig:optimized-deformed-convergence}(a) are quite promising: the estimates using $\pi_\mathrm{opt}$ are slightly better than the ones using $\pi_\mathrm{best}$, which we suspect is due to the crowding of components in the deeper well.

\begin{table}[p]
	\caption{
		Parameters of the model Hamiltonian in Eq.~\eqref{eq:model-system} for the Displaced $\gamma_2$ system.
	}
	\label{tab:displaced-2-parameters}
	\hfill{}
	\begin{tabular}{c S[table-format=1.4]}
		\toprule
		Parameter & {Value} \\
		\midrule
		$E^1$ & 0.0996 \\
		$E^2$ & 0.1996 \\
		$\omega_1$ & 0.02 \\
		$\omega_2$ & 0.04 \\
		\botrule
	\end{tabular}
	\hfill{}
	\begin{tabular}{c S[table-format=1.4]}
		\toprule
		Parameter & {Value} \\
		\midrule
		$\lambda$ & 0.072 \\
		$\gamma$ & 0.04 \\
		\botrule
	\end{tabular}
	\hfill{}
\end{table}

\subsubsection{Displaced $\gamma_6$}

\begin{figure}
	\includegraphics{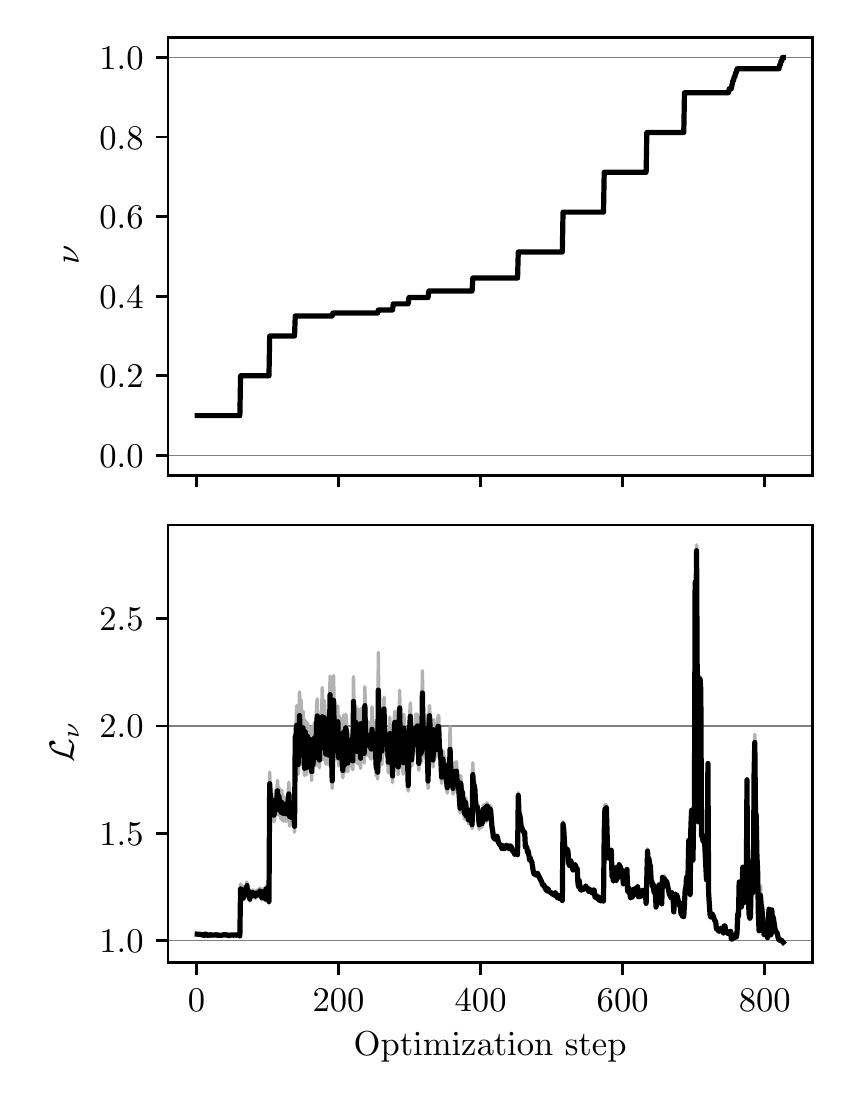}
	\caption{
		The value of $\nu$ (top panel) and the loss function $\mathcal{L}_\nu$ (bottom panel) at each step of the GMD optimization for the Displaced $\gamma_6$ model.
		In the top panel, the horizontal lines indicate the bounds on $\nu$.
		In the bottom panel, the horizontal lines mark the bounds on the desired loss function values, and the uncertainty of the loss function estimates is indicated by the shaded area.
	}
	\label{fig:loss-displaced-6}
\end{figure}

\begin{figure}
	\includegraphics{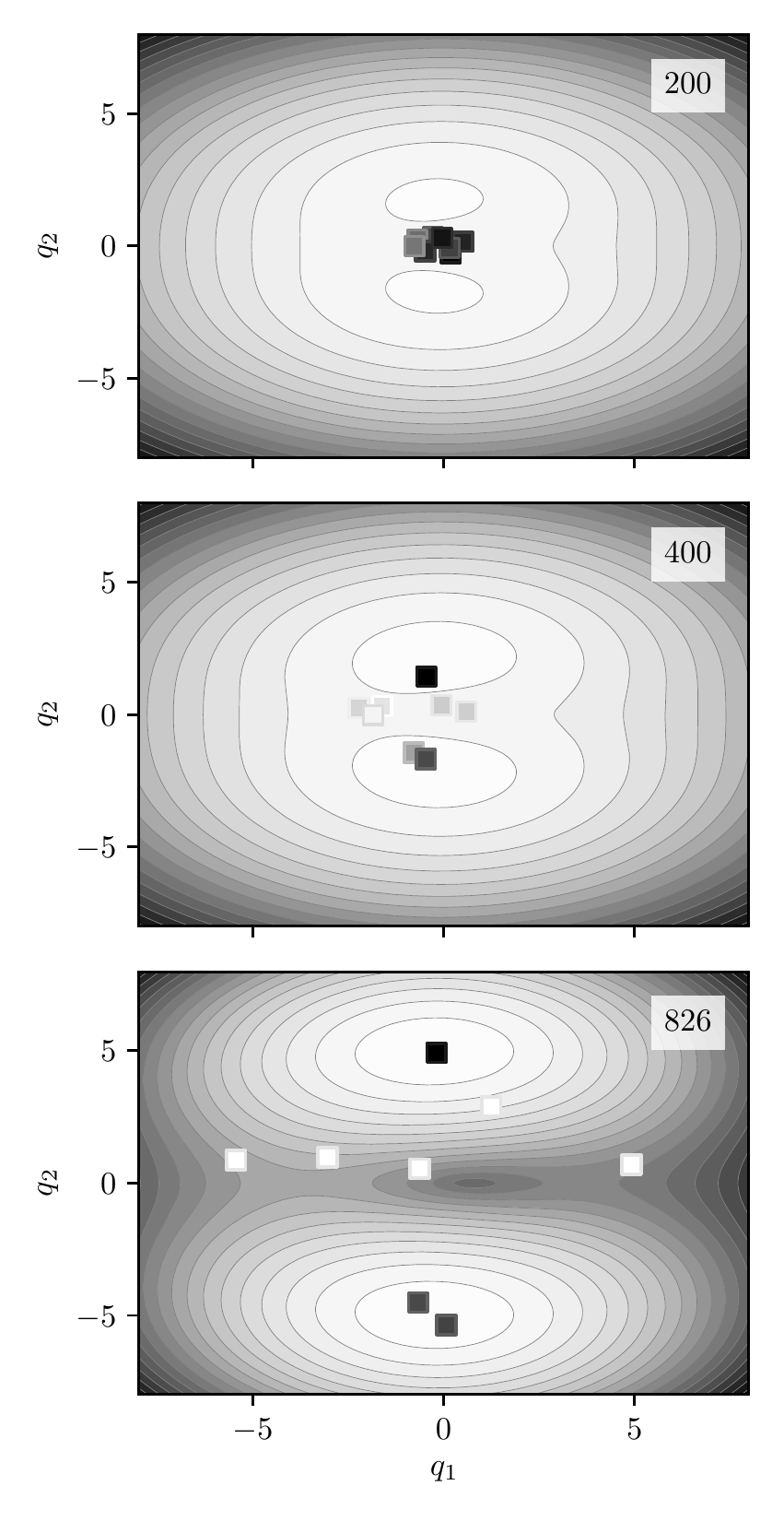}
	\caption{
		Snapshots of the GMD optimization for the Displaced $\gamma_6$ model.
		The number in each panel indicates the optimization step.
		The backgrounds are the ground state potential energy surfaces of $\hat{H}_\nu$.
		The minimum of each GMD component is represented by a square, colored according to the relative component weights.
	}
	\label{fig:pes-displaced-6}
\end{figure}

The first strongly-coupled model system that we consider is the $\gamma_6$ variant of the Displaced model, with the parameters in Tab.~\ref{tab:displaced-6-parameters}, and results in Fig.~\ref{fig:loss-displaced-6} and Fig.~\ref{fig:pes-displaced-6}.
The optimization algorithm struggles when the minima first appear (around steps 100 to 350; top panel of Fig.~\ref{fig:pes-displaced-6}), but it is eventually able to find them.
This difficulty is clear from the large and noisy values of the loss function and the small $\nu$ steps.
However, once the minima are found, the rest of the optimization proceeds fairly well.

Unfortunately, many of the components in the final distribution are of no consequence (bottom panel of Fig.~\ref{fig:pes-displaced-6}).
As they have very low weights, this does not put a damper on the estimation of the partition function seen in Fig.~\ref{fig:optimized-deformed-convergence}(b).
Again, both sampling distributions (in this case $\pi_\mathrm{best}$ is what Ref.~\onlinecite{raymond2018path} calls $\rho_1$) work well, but the optimized distribution $\pi_\mathrm{opt}$ ekes out a tiny decrease in the error bars, possibly due to slightly more thorough coverage of one of the wells.

\begin{table}
	\caption{
		Parameters of the model Hamiltonian in Eq.~\eqref{eq:model-system} for the Displaced $\gamma_6$ system.
	}
	\label{tab:displaced-6-parameters}
	\hfill{}
	\begin{tabular}{c S[table-format=1.4]}
		\toprule
		Parameter & {Value} \\
		\midrule
		$E^1$ & 0.0996 \\
		$E^2$ & 0.1996 \\
		$\omega_1$ & 0.02 \\
		$\omega_2$ & 0.04 \\
		\botrule
	\end{tabular}
	\hfill{}
	\begin{tabular}{c S[table-format=1.4]}
		\toprule
		Parameter & {Value} \\
		\midrule
		$\lambda$ & 0.072 \\
		$\gamma$ & 0.2 \\
		\botrule
	\end{tabular}
	\hfill{}
\end{table}

\subsubsection{Jahn--Teller $\lambda_6$}

\begin{figure}
	\includegraphics{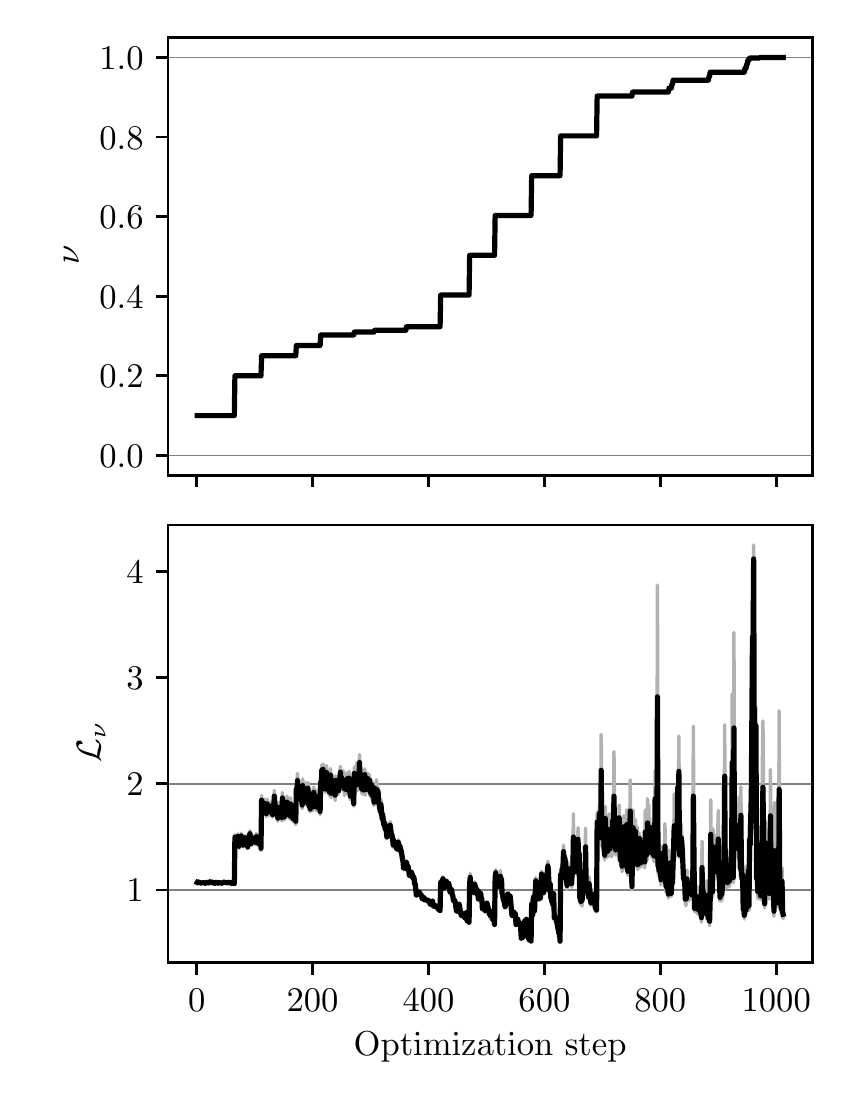}
	\caption{
		The value of $\nu$ (top panel) and the loss function $\mathcal{L}_\nu$ (bottom panel) at each step of the GMD optimization for the Jahn--Teller $\lambda_6$ model.
		In the top panel, the horizontal lines indicate the bounds on $\nu$.
		In the bottom panel, the horizontal lines mark the bounds on the desired loss function values, and the uncertainty of the loss function estimates is indicated by the shaded area.
	}
	\label{fig:loss-jahnteller-6}
\end{figure}

\begin{figure}
	\includegraphics{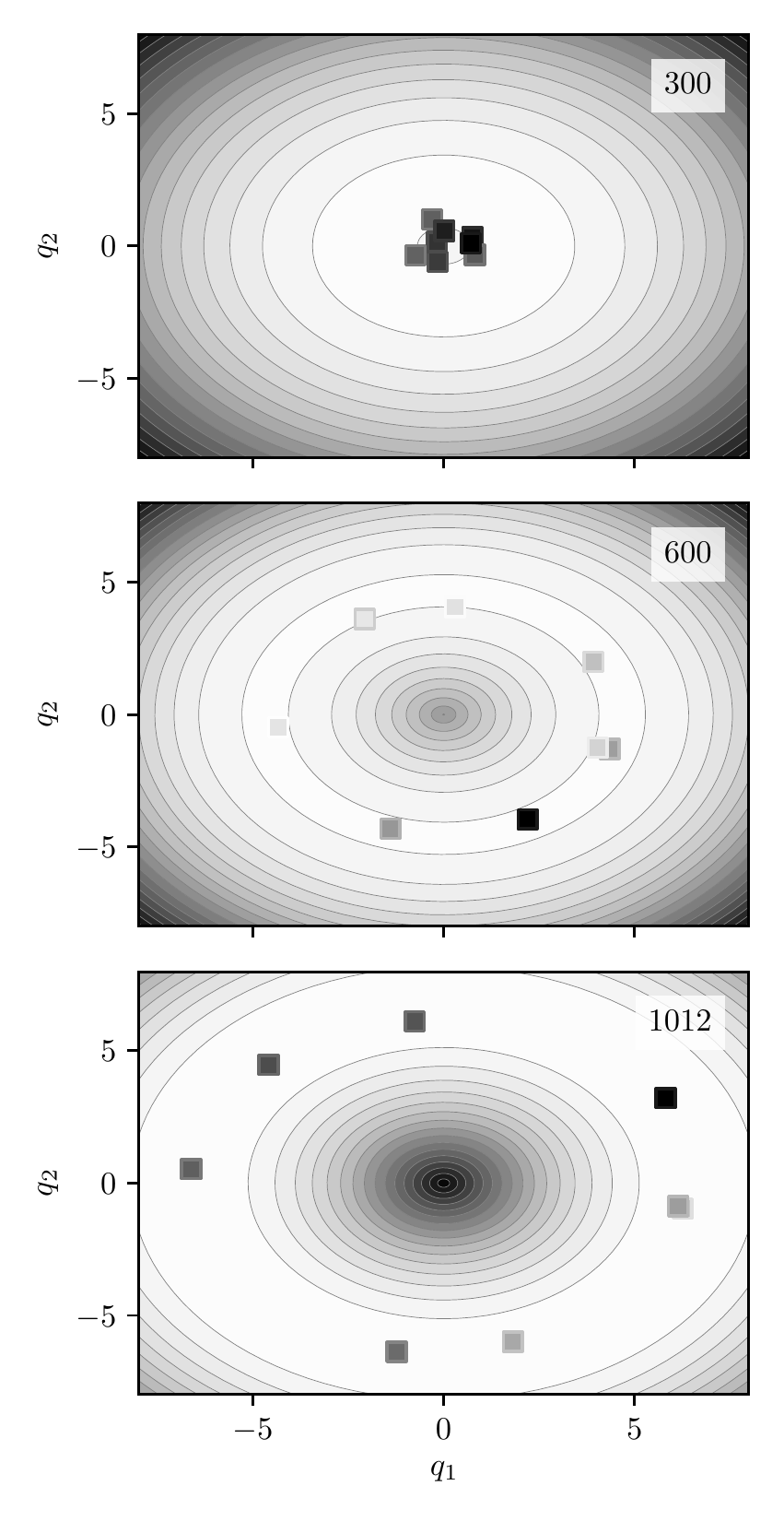}
	\caption{
		Snapshots of the GMD optimization for the Jahn--Teller $\lambda_6$ model.
		The number in each panel indicates the optimization step.
		The backgrounds are the ground state potential energy surfaces of $\hat{H}_\nu$.
		The minimum of each GMD component is represented by a square, colored according to the relative component weights.
	}
	\label{fig:pes-jahnteller-6}
\end{figure}

As the other strongly-coupled model system we use the $\lambda_6$ variant of the Jahn--Teller model from Ref.~\onlinecite{raymond2018path}, whose parameters are given in Tab.~\ref{tab:jahn-teller-6-parameters}.
The results can be found in Fig.~\ref{fig:loss-jahnteller-6} and Fig.~\ref{fig:pes-jahnteller-6}.
The early part of the optimization is a bit uncertain, as the components search for the circular well (top panel of Fig.~\ref{fig:pes-jahnteller-6}).
Once they fall into the well around step 350, the situation temporarily improves.
However, as the symmetric well grows, more regions become neglected and loss function estimation becomes more difficult.
Additionally, there are significant fluctuations in the components weights; at times, only a single component will have most of the weight (middle panel of Fig.~\ref{fig:pes-jahnteller-6}).
Still, by the end of the optimization (bottom panel of Fig.~\ref{fig:pes-jahnteller-6}), the components are reasonably well spread out with somewhat even weights, which attests to the importance of the penalty terms in Eq.~\eqref{eq:loss-deform} for avoiding clustering of components and domination by a single component.

\begin{table}
	\caption{
		Parameters of the model Hamiltonian in Eq.~\eqref{eq:model-system} for the Jahn--Teller $\lambda_6$ system.
	}
	\label{tab:jahn-teller-6-parameters}
	\hfill{}
	\begin{tabular}{c S[table-format=1.5]}
		\toprule
		Parameter & {Value} \\
		\midrule
		$E^1$ & 0.63135 \\
		$E^2$ & 0.63135 \\
		$\omega_1$ & 0.03 \\
		$\omega_2$ & 0.03 \\
		\botrule
	\end{tabular}
	\hfill{}
	\begin{tabular}{c S[table-format=1.5]}
		\toprule
		Parameter & {Value} \\
		\midrule
		$\lambda$ & 0.2 \\
		$\gamma$ & 0.2 \\
		\botrule
	\end{tabular}
	\hfill{}
\end{table}

Likely owing to the uneven distribution of the components in the well, the partition function estimation in Fig.~\ref{fig:optimized-deformed-convergence}(c) does not perform as well with the optimized distribution $\pi_\mathrm{opt}$ as it does with $\pi_\mathrm{best}$ (in this case, $\rho_2$ of Ref.~\onlinecite{raymond2018path}, which also has 8 components).
This could conceivably be remedied by including more components in the GMD or adjusting the $\epsilon$ and $C_1$ parameters responsible for the repulsion penalty term of the loss function.

\section{Conclusions}
\label{sec:conclusions}

We have demonstrated four methods to enhance partition function estimation for vibronic Hamiltonians: deterministic component selection in GMD sampling when using MC (Sec.~\ref{sec:methods-component}, Sec.~\ref{sec:results-component}); RqMC as a substitute for MC in GMD sampling (Sec.~\ref{sec:methods-quasi}, Sec.~\ref{sec:results-quasi}); and optimization of the GMD parameters without (Sec.~\ref{sec:methods-optimization}, Sec.~\ref{sec:results-optimization}) and with deformation (Sec.~\ref{sec:methods-optimization-deformation}, Sec.~\ref{sec:results-optimization-deformation}).
Each of these was shown to improve the scheme described in Ref.~\onlinecite{raymond2018path}.

We have found that choosing the components of a GMD in an MC calculation deterministically as opposed to stochastically can slightly reduce the standard error of the mean with a fixed number of samples (or, conversely, reduce the number of samples for a fixed standard error).
Additionally, it can diminish the impact of extreme outliers.
Thus, we recommend the use of deterministic component selection for MC partition function calculations of vibronic models, and for GMD sampling in general.

However, when possible, RqMC should be used in place of MC.
We have shown that employing quasi-random numbers in addition to pseudo-random numbers can help improve the rate at which the stochastic error decreases with the number of samples, making RqMC the more efficient choice.
In subsequent studies, the impact of the $N_\mathrm{S}$ and $N_\mathrm{boot}$ parameters on the quality of the error estimate should be determined.

Finally, the optimization of GMD parameters was observed to improve the sampling efficiency when compared to the simplest choice of sampling distribution, and sometimes even when compared to a hand-crafted sampling distribution.
For the version without deformation of the target distribution, we have presented an example of a distribution which is significantly improved in just 256 steps.
Because of the way the sampling GMD is constructed from Hamiltonian parameters, even though the optimization was carried out at $P = 64$ beads, the distribution obtained for $P = 1024$ functions very well.

The iterative algorithm to optimize GMD parameters with deformation of the target distribution was seen to perform rather well for two of the three systems, but only tolerably for the third.
It was able to find the regions of high density for each model system, resulting in very well-behaved sampling at several values of $\tau$, although the highly symmetric Jahn--Teller $\lambda_6$ system was shown to give the method some trouble.
Because the algorithm parameter values were the same across all the model systems, this appears to be a fairly robust and general method, and we are interested in seeing it applied to realistic molecular systems.

Several upgrades to the optimization are possible which may improve its efficiency, and which we hope to attempt in the future.
The first is to use different magnitudes for the different components of the SPSA perturbation vector $\vec{\Delta}$, allowing the $E^{(0)}$ parameters to evolve at a different rate than the $E^{(1)}$ parameters.
Another is to allow low-weight components to be culled and replaced by clones of high-weight components.
Finally, if the loss function is seen to be decreasing rapidly, it may be beneficial to continue beyond $N_\mathrm{SPSA}$ optimization steps before moving on to the next value of $\nu$.

\begin{acknowledgments}
P.-N.R. (RGPIN-2016-04403) and M.N. (RGPIN-2018-04187) acknowledge the Natural Sciences and Engineering Research Council of Canada (NSERC).
P.-N.R. acknowledges the Ontario Ministry of Research and Innovation (MRI), the Canada Research Chair program (950-231024), the Canada Foundation for Innovation (CFI) (project No. 35232), and the Canada First Research Excellence Fund (CFREF).
\end{acknowledgments}

\clearpage

\appendix

\section{Demonstration of quasi-Monte Carlo}
\label{sec:qmc-demo}

\begin{figure}
	\includegraphics{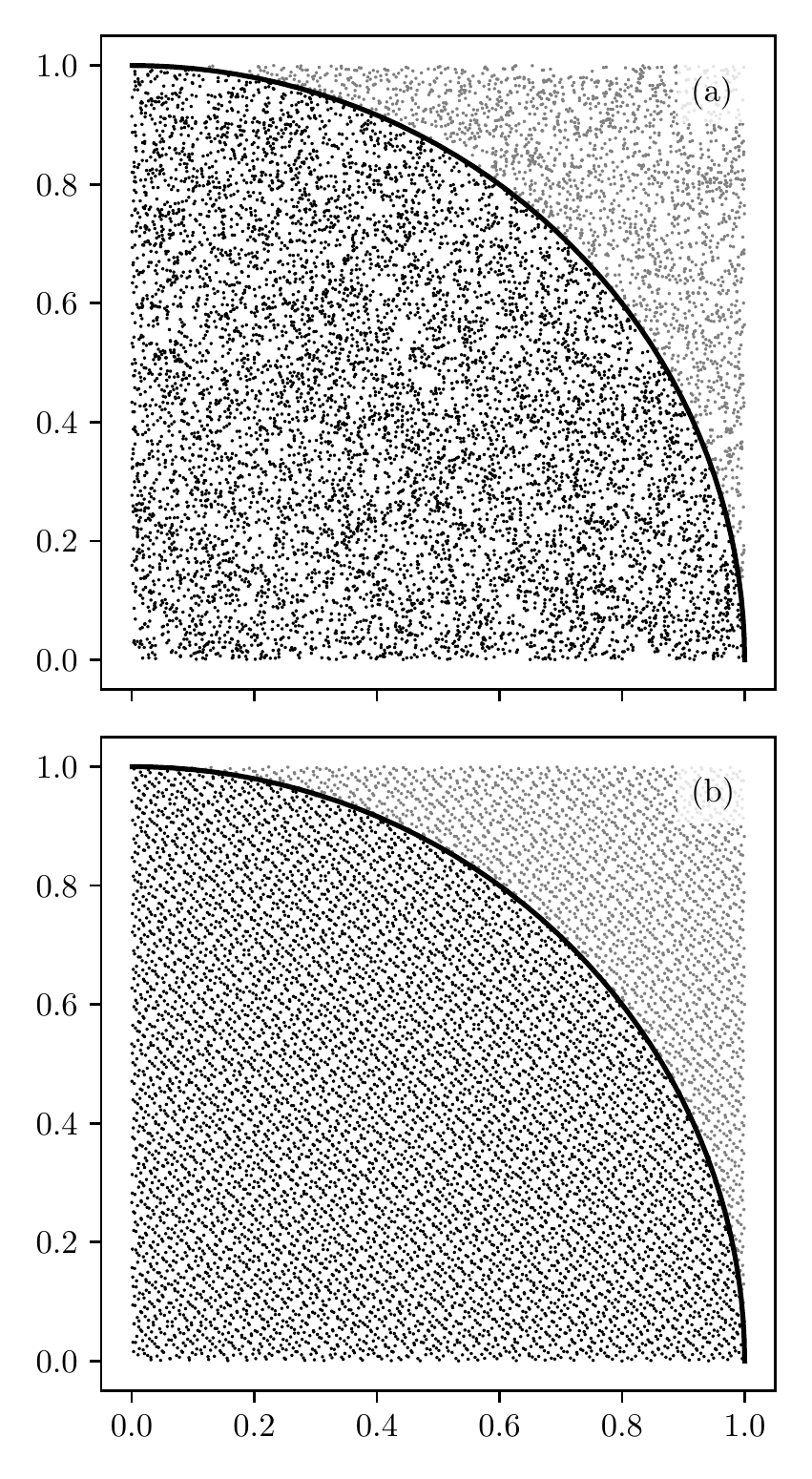}
	\caption{
		Distribution of \num{e4} points sampled (a) pseudo-randomly using a Mersenne Twister RNG and (b) quasi-randomly using a Sobol sequence.
		The curve is a segment of the unit quarter-circle centered at the origin; points inside it are darker, and those outside are lighter.
	}
	\label{fig:pi-sample}
\end{figure}

At its core, quasi-Monte Carlo (qMC) is just plain Monte Carlo (MC), but with the pseudo-random number generator (RNG) replaced by a low-discrepancy sequence (LDS).
Informally, the discrepancy of a sequence is its maximum local deviation from a uniform density.\cite{niederreiter1978quasi}
LDSs attempt to minimize the discrepancy of the generated points and therefore cover space more evenly than RNGs.\cite{caflisch1998monte}
The hope is that this can help reduce the stochastic error when using a statistical estimator.

The canonical introductory MC problem is that of estimating $\pi$ (the ratio of a circle's circumference to its diameter) by sampling uniformly from a square of unit side length, and counting the fraction $F$ of points lying within one unit of one of the corners.
Since the ratio of areas is $\pi / 4$, $4 F$ should tend to $\pi$.
In this section, we compare the behavior of MC to qMC for this elementary problem.
For the former, we use the Mersenne Twister, an extremely popular RNG;\cite{matsumoto1998mersenne} for the latter, we use a two-dimensional Sobol sequence.\cite{sobol1967distribution}

\begin{figure}
	\includegraphics{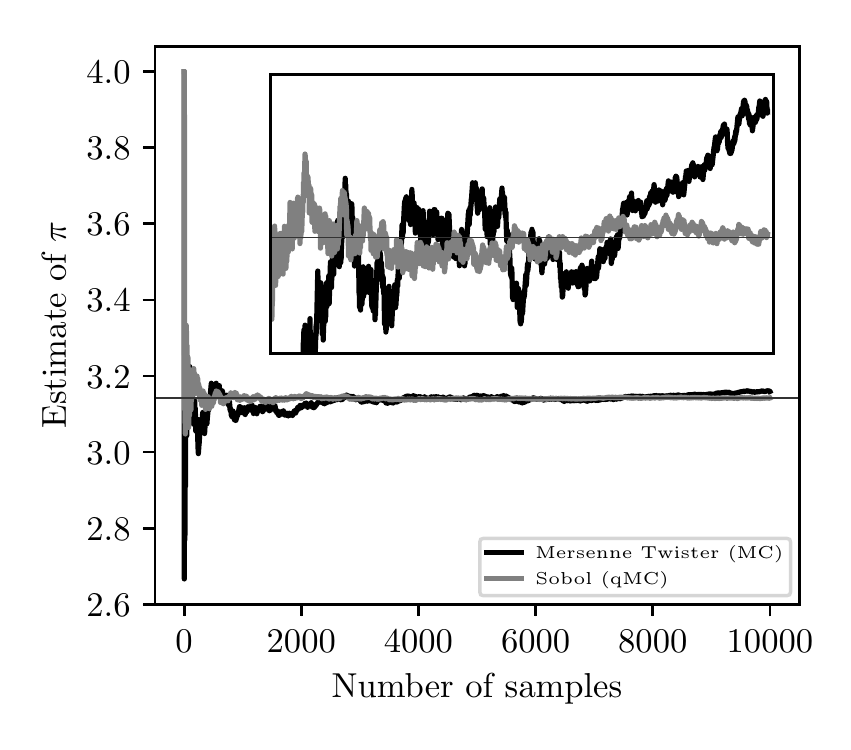}
	\caption{
		Convergence of MC and qMC methods for the estimation of $\pi$.
		The exact value is shown with a horizontal line.
		The inset shows the data directly below it.
	}
	\label{fig:pi-convergence}
\end{figure}

As seen in Fig.~\ref{fig:pi-sample}(a), the RNG creates a typically irregular pattern, with clumps and empty spaces.
On the other hand, in Fig.~\ref{fig:pi-sample}(b), the quasi-random points are extremely uniform.
It is this regularity that results in the faster convergence of qMC shown in Fig.~\ref{fig:pi-convergence}.
Although qMC is not a panacea for numerical integration, this example demonstrates that it can have improved performance over plain MC.

\section{Gaussian sampling in quasi-Monte Carlo}
\label{sec:qmc-gaussian}

\begin{figure*}
	\includegraphics{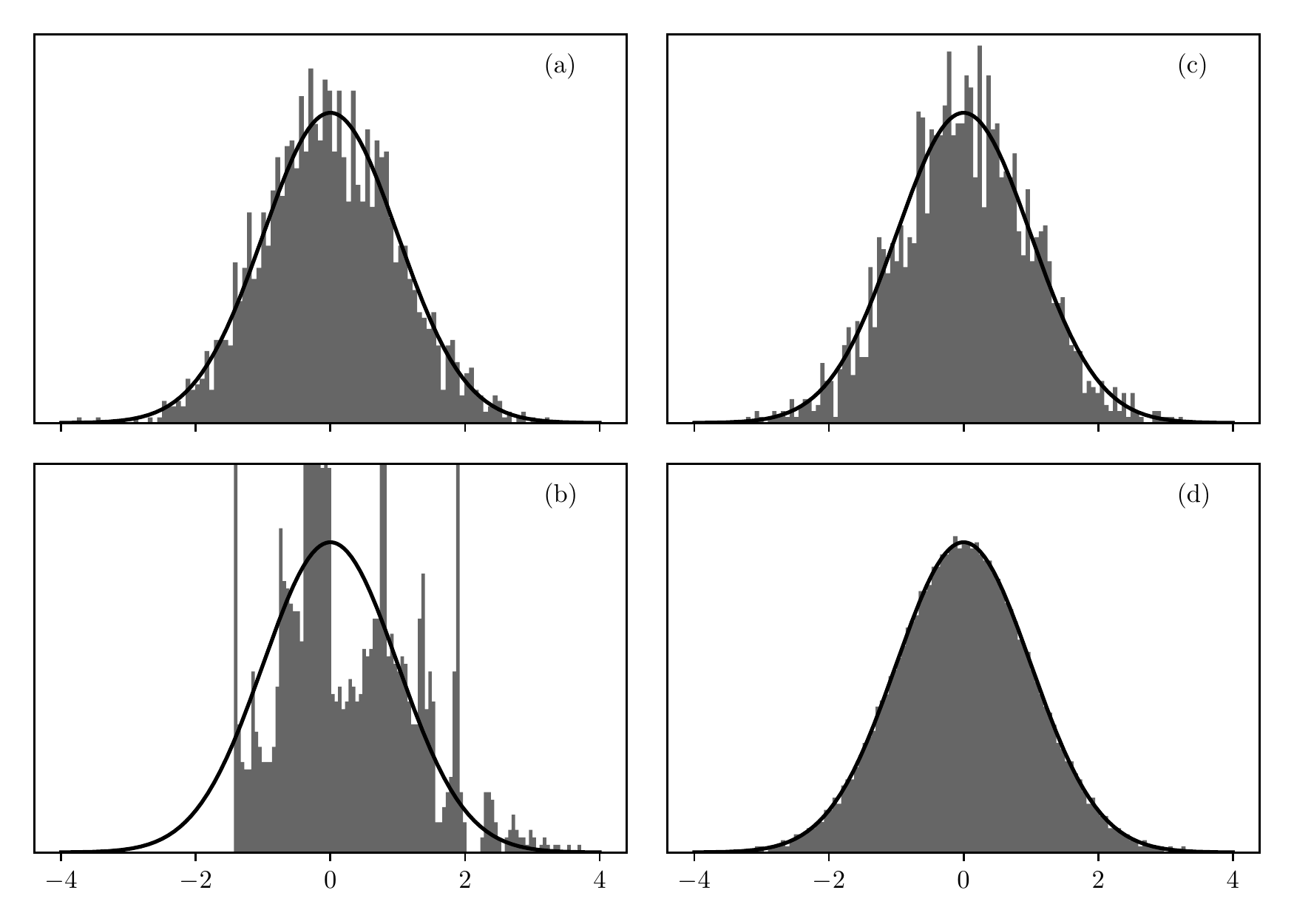}
	\caption{
		Histograms comparing pseudo-random and quasi-random Gaussian sampling methods: (a) Box--Muller with Mersenne Twister RNG; (b) Box--Muller with Sobol LDS; (c) cdf inversion with Mersenne Twister RNG; (d) cdf inversion with Sobol LDS.
		Each histogram is generated from \num{2000} samples grouped into \num{100} bins.
		In (c), some bins are cut off at the top to show more detail elsewhere.
		The true pdf is drawn as a solid curve.
	}
	\label{fig:rng}
\end{figure*}

One method for sampling from an arbitrary one-dimensional probability distribution function (pdf) uses the inverse of the corresponding cumulative distribution function (cdf).
If uniformly distributed samples are input into the inverse cdf, the outputs are distributed according to the desired pdf.
Unfortunately, this is often computationally taxing, so specialized methods are preferred.
For the case of Gaussian random variables, two common alternatives are Box--Muller and ziggurat.\cite{box1958note,marsaglia2000ziggurat}

While these methods perform well with pseudo-random numbers, they destroy the low-discrepancy nature of quasi-random numbers.
Compare, for example, the histogram obtained from sampling a standard Gaussian with the Box--Muller method and the Mersenne Twister pseudo-random number generator (RNG) in Fig.~\ref{fig:rng}(a) and the same method applied to a Sobol low-discrepancy sequence (LDS) in Fig.~\ref{fig:rng}(b).
The latter does an extremely poor job of describing the true pdf.

Despite its cost, the preferred method in this circumstance turns out to be straightforward cdf inversion.\cite{joy1996quasi,brown2013self}
Application of this method is shown in Fig.~\ref{fig:rng}(c) using the Mersenne Twister RNG and in Fig.~\ref{fig:rng}(d) using a Sobol LDS.
It is evident that among the four options presented in Fig.~\ref{fig:rng}, cdf inversion of an LDS results in by far the smoothest histogram.
Thus, we have chosen this approach as the basis of low-discrepancy Gaussian sampling in the present work.

\begin{figure}
	\includegraphics{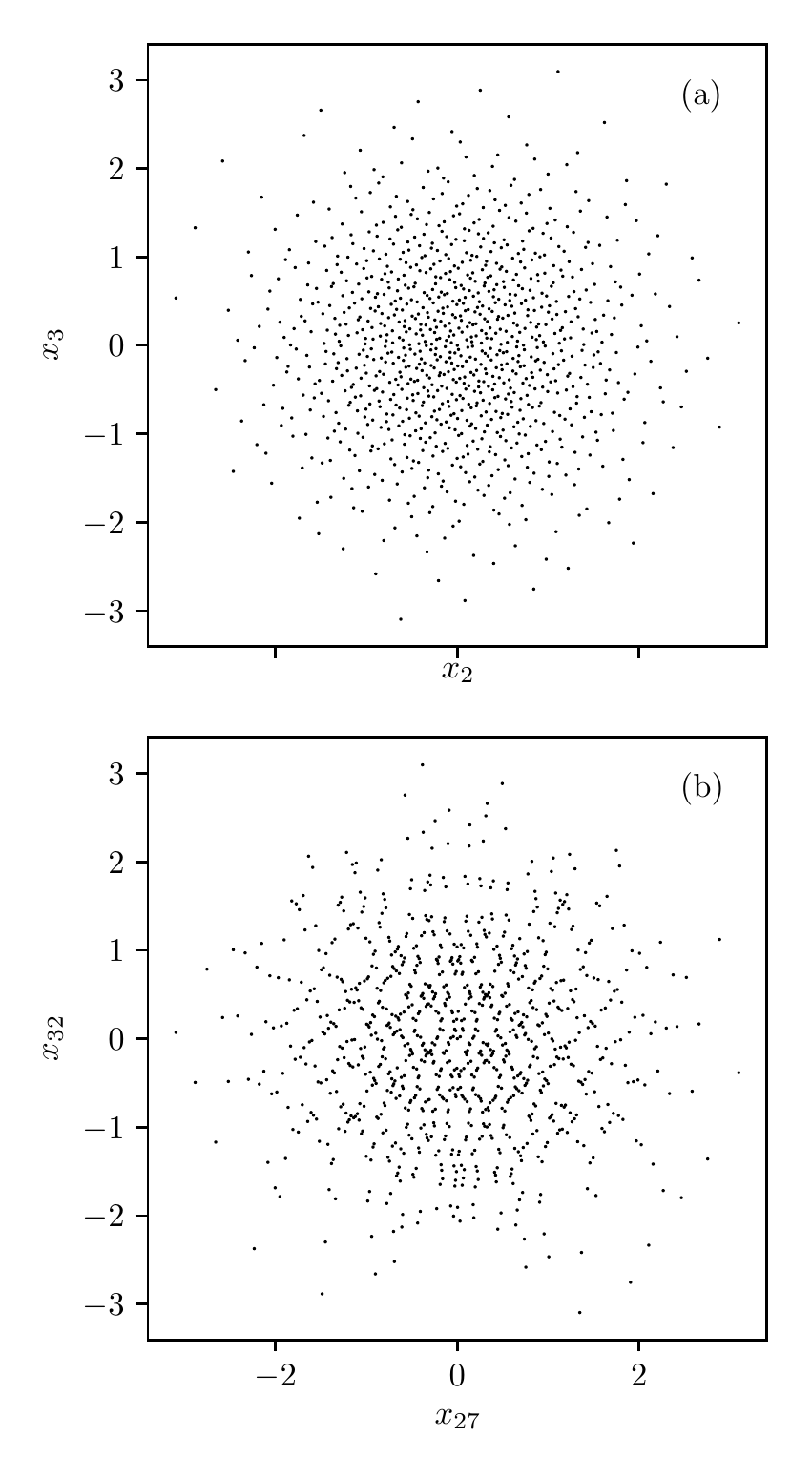}
	\caption{
		Cuts along dimensions (a) 2 and 3 and (b) 27 and 32 of a 32-dimensional Gaussian LDS point set containing \num{1000} points.
	}
	\label{fig:sobol-coupling}
\end{figure}

Nevertheless, even cdf inversion is imperfect with a many-dimensional LDS.
Because the higher dimensions of a Sobol sequence tend to be more strongly correlated before the inversion,\cite{caflisch1998monte} these correlations are carried through to the generated points.
In Fig.~\ref{fig:sobol-coupling}, we show two-dimensional cuts of a 32-dimensional point set created from a 32-dimensional LDS using cdf inversion separately for each dimension.
While there are no obvious correlations present in Fig.~\ref{fig:sobol-coupling}(a) between the second and third dimensions, we see some glaring patterns in Fig.~\ref{fig:sobol-coupling}(b), where dimensions 27 and 32 are shown.
However, these correlations do not prevent us from using quasi-random numbers as a smoother alternative to pseudo-random numbers for multidimensional Gaussian sampling.

\section{Example of randomized quasi-Monte Carlo}
\label{sec:rqmc}

The flexibility allowed in randomized quasi-Monte Carlo (RqMC) by changing the number of sequences $N_\mathrm{S}$ creates a tension between pure qMC ($N_\mathrm{S} = 1$), which lacks the ability to accurately estimate error bars, and pure Monte Carlo ($N_\mathrm{S} = N_\mathrm{MC}$), which misses out on the low-discrepancy smoothness.
In this section, we demonstrate the practical implications of this balance by estimating the 64-dimensional integral
\begin{align}
	\label{eq:rqmc-example-integral}
	I
	&= \frac{1}{(2\pi)^{32}} \int\! \dd{x_1} \cdots \int\! \dd{x_{64}} \,
			e^{-\sum_{j=1}^{64} \frac{x_j^2}{2}} \sum_{j=1}^{64} x_j^2.
\end{align}
Factoring and evaluating these integrals individually reveals the exact solution $I = 64$.

To compute $I$ using Monte Carlo (MC), we sample $N_\mathrm{MC} = 2^{14}$ times from a collection of 64 uncoupled standard Gaussian distributions.
For RqMC, we instead obtain $N_\mathrm{qMC} = N_\mathrm{MC} / N_\mathrm{S}$ samples from each of $N_\mathrm{S}$ 64-dimensional Sobol sequences, then transform them using cdf inversion, and combine them into a single point with error bars.
In either case, we use the estimator
\begin{align}
	\sum_{j=1}^{64} x_j^2
\end{align}
and the total number of sampled points is always the same.

\begin{figure}
	\includegraphics{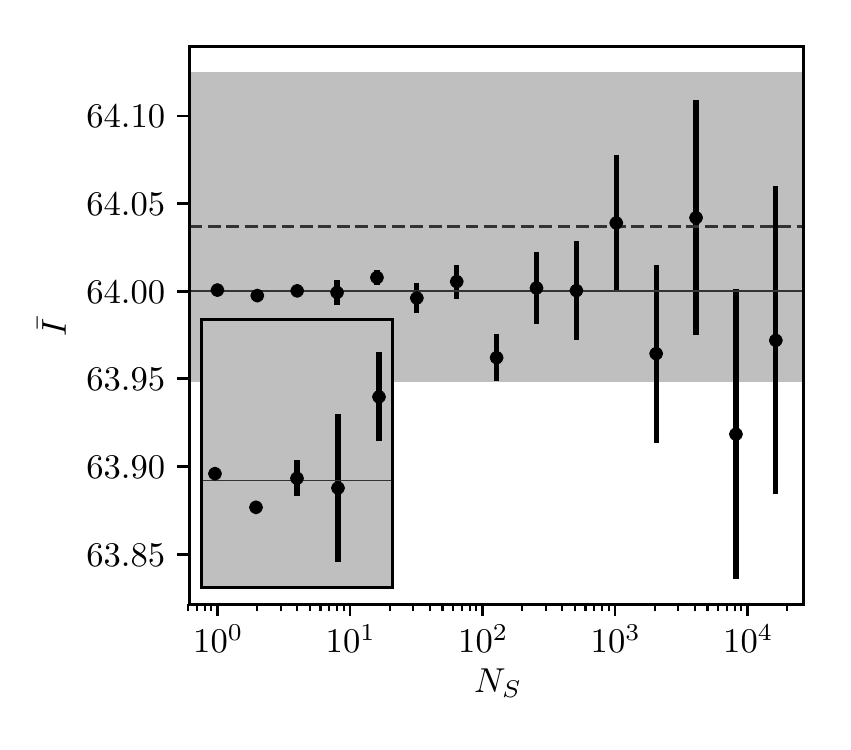}
	\caption{
		Estimation of $I$ in Eq.~\eqref{eq:rqmc-example-integral} using RqMC.
		The exact value is shown with a solid horizontal line.
		The MC result is shown with a dashed horizontal line and the shaded area indicates the extent of the MC error bars.
		The inset shows the data directly above it.
	}
	\label{fig:rqmc-example}
\end{figure}

The results in Fig.~\ref{fig:rqmc-example} make it clear that for such a well-behaved integrand, RqMC does not require all that many sequences in order to generate faithful error bars.
Additionally, the error bars for $N_\mathrm{S} \ll N_\mathrm{MC}$ are substantially smaller than the MC ones, suggesting improved performance.
When $N_\mathrm{S} \approx N_\mathrm{MC}$, the RqMC error bars are of a comparable size to their MC counterpart, which confirms that RqMC matches MC in this limit.

\section{Variance of linear combination}
\label{sec:variance-estimator}

Consider a collection of $B$ uncorrelated random variables $\{ r^b \}_{b=1}^B$, from which we form a new random variable
\begin{align}
	r
	&= \sum_{b=1}^B w^b r^b
\end{align}
using the constant coefficients $w^b$.
If each $r^b$ has population variance
\begin{align}
	\sigma^2_{r^b}
	&= \ev{\left( r^b - \ev*{r^b} \right)^2}
	= \ev*{(r^b)^2} - \ev*{r^b}^2,
\end{align}
then
\begin{subequations}
\begin{align}
	\sigma^2_r
	&= \ev{\left( r - \ev*{r} \right)^2} \\
	&= \sum_{b=1}^B \sum_{b'=1}^B w^b w^{b'} \ev{(r^b - \ev*{r^b}) (r^{b'} - \ev*{r^{b'}})}.
\end{align}
\end{subequations}
For $b \ne b'$, $r^b$ and $r^{b'}$ are uncorrelated, so
\begin{align}
	\ev{(r^b - \ev*{r^b}) (r^{b'} - \ev*{r^{b'}})}
	&= 0.
\end{align}
Thus,
\begin{align}
	\sigma^2_r
	&= \sum_{b=1}^B (w^b)^2 \ev{\left( r^b - \ev*{r^b} \right)^2}
	= \sum_{b=1}^B (w^b)^2 \sigma^2_{r^b}.
\end{align}

Suppose that $s^2_{r^b}$ is an unbiased estimator of $\sigma^2_{r^b}$, meaning that
\begin{align}
	\ev{s^2_{r^b}}
	&= \sigma^2_{r^b}.
\end{align}
Then
\begin{align}
	s^2_r
	&= \sum_{b=1}^B (w^b)^2 s^2_{r^b}
\end{align}
is an unbiased estimator of $\sigma^2_r$, since
\begin{align}
	\ev{s^2_r}
	&= \sum_{b=1}^B (w^b)^2 \ev{s^2_{r^b}}
	= \sum_{b=1}^B (w^b)^2 \sigma^2_{r^b}
	= \sigma^2_r.
\end{align}

\section{Example of negative $g(\vec{R})/Z$}
\label{sec:negative-g}

It was noted in Ref.~\onlinecite{raymond2018path} that $g(\vec{R})$, as defined in Eq.~\eqref{eq:gR}, may take on negative values.
If $g(\vec{R}) > 0$ for some $\vec{R}$, but $g(\vec{R}) < 0$ for others, $g(\vec{R})$ cannot always agree in sign with any constant real number $Z$, so $g(\vec{R})/Z$ must be negative at some point and is not a pdf.
We give a simple example of this using the Hamiltonian
\begin{align}
	\label{eq:negative-g-model-system}
	\hat{H}
	&= \begin{pmatrix}
				\hat{h}_\mathrm{o} & 0 \\
				0 & \hat{h}_\mathrm{o} \\
			\end{pmatrix}
		+ \begin{pmatrix}
				\gamma_{1,1} & \gamma_{1,2} \\
				\gamma_{1,2} & \gamma_{2,2} \\
			\end{pmatrix} \hat{q}_1,
\end{align}
with the parameters in Tab.~\ref{tab:negative-g-parameters}.
At an inverse temperature of $\beta = 38.7$, with $P = 3$ beads located at $\vec{R} = (1, 2, 0)$, we obtain
\begin{align}
	g(\vec{R})
	&\approx \num{0.0031165}.
\end{align}
However, at $\vec{R} = (1, 2, -0.3)$, we instead find
\begin{align}
	g(\vec{R})
	&\approx \num{-0.010741}.
\end{align}

\begin{table}
	\caption{
		Parameters of the model Hamiltonian in Eq.~\eqref{eq:negative-g-model-system}.
	}
	\label{tab:negative-g-parameters}
	\hfill{}
	\begin{tabular}{c S[table-format=1.3]}
		\toprule
		Parameter & {Value} \\
		\midrule
		$\omega_1$ & 0.1 \\
		$\gamma_{1,1}$ & 1.0 \\
		\botrule
	\end{tabular}
	\hfill{}
	\begin{tabular}{c S[table-format=1.3]}
		\toprule
		Parameter & {Value} \\
		\midrule
		$\gamma_{1,2}$ & 0.1 \\
		$\gamma_{2,2}$ & 0.01 \\
		\botrule
	\end{tabular}
	\hfill{}
\end{table}

\section{Explanation of $\nu$ step size selection algorithm}
\label{sec:step-nu}

The function \textsc{step\_nu} for choosing the step size $\Delta \nu$ in Sec.~\ref{sec:methods-optimization-deformation} is given in Alg.~\ref{alg:step-nu}, and is based on a number of heuristics.
The core assumption is that (all else being fixed) the loss function grows monotonically with the step size.
We justify this by pointing out that the relative entropy term which appears in the loss function quantifies the difference between the GMD used for sampling and the target distribution, and this difference should grow with increasing deformation of the target distribution.

We use this assumption to tune the loss function to be between 1 and 2 by varying $\Delta \nu$ with a simple binary search.
If the loss function is too low (and we would not be optimizing very much this iteration), the lower bound $\Delta \nu_\mathrm{min}$ is set to the current value of $\Delta \nu$; if it is too high (and we might be approaching a regime  where the sampling is not trustworthy), the upper bound $\Delta \nu_\mathrm{max}$ is set to the current value of $\Delta \nu$.
After each update, the new trial value of $\Delta \nu$ is set to be the average of the lower and upper bounds.
The values of 1 and 2 for the loss function bounds, as well as the number of binary search iterations, are entirely ad hoc, but appear to be reasonable for the systems considered in this work.

We start the process by optimistically guessing that the current step should be twice as large as the previous step.
We also take steps that are no smaller than $\num{e-3}$ so that the optimization doesn't grind to a halt in difficult regions, and no larger than $\num{e-1}$ (or $\num{e-2}$ close to the end) to ensure that we don't skip over any details.

\end{document}